\documentclass[11pt]{article}
\usepackage{epsfig,sint,macros,cite,mathbbol}

\usepackage[english]{babel}
\usepackage{simplewick}

\begin{document}

\begin{titlepage}
\begin{flushright}
\hfill CERN-PH-TH/2006-259\\
\hfill \ \ \ \ \ \ \   IFIC/06-45       
\end{flushright}

\vskip 0.5 cm
\begin{center}
  {\Large\bf 
  Spontaneous chiral symmetry breaking in QCD: \\
  a finite-size scaling study on the lattice\\[0.5ex] } 
\end{center}
\vskip 0.5 cm
\begin{center}
{\large 
     Leonardo Giusti$^{\scriptscriptstyle a}$ and
     Silvia Necco$^{\scriptscriptstyle b}$
}
\vskip 0.5cm
$^{\scriptstyle a}$
CERN, Physics Department, TH Division, CH-1211 Geneva 23, Switzerland
\vskip 1.5ex
$^{\scriptstyle b}$
Instituto de F\'isica Corpuscular, CSIC-Universitat de Val\`encia\\
Apartado de Correos 22085, E-46071 Valencia, Spain
\vskip 1.0cm
{\bf Abstract}
\vskip 0.35ex
\end{center}

\noindent
Spontaneous chiral symmetry breaking in QCD with massless quarks 
at infinite volume can be seen in a finite box by studying, for instance, 
the dependence of the chiral condensate from the volume and the quark mass.
We perform a feasibility study of this program by computing the quark condensate on 
the lattice in the quenched approximation of QCD at small quark masses. 
We carry out simulations in various topological sectors of the theory 
at several volumes, quark masses and lattice spacings by employing fermions with an 
exact chiral symmetry, and we focus on observables which 
are infrared stable and free from mass-dependent ultraviolet divergences. 
The numerical calculation is carried out with an exact variance-reduction 
technique, which is designed to be particularly efficient when spontaneous 
symmetry breaking is at work in generating a few very small low-lying 
eigenvalues of the Dirac operator. The finite-size scaling behaviour of the
condensate in the topological sectors considered agrees, within our statistical 
accuracy, with the expectations of the chiral effective theory. Close to the chiral 
limit we observe a detailed agreement with the first Leutwyler--Smilga sum rule. By 
comparing the mass, the volume and the topology dependence of our results 
with the predictions of the chiral effective theory, we extract the corresponding 
low-energy constant.
\vfill

\eject

\vfill

\eject

\end{titlepage}

\section{Introduction}
Spontaneous symmetry breaking plays a central r\^ole in 
our understanding of the strong interactions. In QCD 
with (a small number of) $N_f$ light flavors the standard 
expectation is that the $SU(N_f)_L \times SU(N_f)_R$ chiral 
symmetry group breaks spontaneously to $SU(N_f)_{L+R}$ with the formation 
of a non-zero quark condensate. 
There are many evidences that support this 
picture~\cite{Weinberg:1978kz,Gasser:1983yg,
DelDebbio:2006cn,Boucaud:2007uk,Giusti:1998wy,Blum:2000kn,
Hernandez:1999cu,Giusti:2001pk,Hernandez:2001hq,DeGrand:2001ie,
Hasenfratz:2002rp,Giusti:2003gf,Wennekers:2005wa,Bietenholz:2006fj}, but 
a conclusive study with a reliable determination of the 
condensate is still missing\footnote{For a recent review of 
the experimental implications of spontaneous symmetry 
breaking in QCD see Ref.~\cite{Colangelo:2001df} and references 
therein.}. 

The lattice formulation of gauge theories is at present
the only approach where QCD can be defined non-perturbatively.
Computations in this regime are performed by 
numerical simulations, and therefore limited to be at 
finite volume and lattice spacing. The spontaneous 
formation of a quark condensate $\Sigma$ at infinite volume
can be detected in a finite box by studying the 
properties of the long-range correlations left in the 
system, a technique widely used in statistical 
mechanics~\cite{Zinn-Justin:2002ru}. In QCD the simplest observable 
to consider is maybe the chiral condensate generated 
by adding a small quark mass $m$ to the action. This 
explicit breaking of chiral symmetry gives also rise to 
additive ultraviolet divergences which make the 
definition of the condensate ambiguous. The crucial observation 
here is that the ultraviolet divergences are independent on 
the volume $V$ and, in a regularization that preserve chiral symmetry, 
their magnitude is suppressed by a factor $1/V$ with respect to  
the contributions from the long-range correlations~\cite{Leutwyler:1992yt}.
At any given lattice spacing the value of $\Sigma$ can then be
extracted with arbitrary precision from a finite-size scaling 
study of the condensate if the volumes are large enough.

In a finite box it is also interesting to consider QCD 
in a fixed topological sector, i.e. with the functional 
integral restricted to gauge-field configurations with a 
given topological charge. Recent progress in the 
understanding of topology on the 
lattice~\cite{Giusti:2001xh,Giusti:2004qd,Luscher:2004fu}
allows to show that the chiral condensate 
has the same ultraviolet divergences in the full theory
and at fixed topology. It can be made 
ultraviolet finite with the very same renormalization 
constants and subtraction coefficients. On the other hand 
the chiral effective theory makes definite predictions, with no 
extra free parameters, for the topology dependence of 
the infrared contribution to 
condensate~\cite{Leutwyler:1992yt,Osborn:1998qb,Damgaard:1998xy},
predictions which can be directly tested against lattice QCD results.

Despite of many efforts dedicated to implement this program in 
the lattice community~\cite{Hernandez:1999cu,DeGrand:2001ie,Hasenfratz:2002rp}, 
a reliable determination of the QCD chiral condensate is still missing. 
The main reasons being the limitations 
in our ability of going beyond the quenched approximation in the numerical
simulations of QCD, and the large numerical instabilities in the commonly 
used estimators of the chiral condensate at finite volume~\cite{Hasenfratz:2002rp}.
The first problem is being solved over the last few years thanks to the 
development of new algorithms for simulating dynamical 
fermions~\cite{Luscher:2004rx,Urbach:2005ji}. The implementation of some of 
these new ideas to simulate dynamical Ginsparg--Wilson fermions 
is progressing quite fast. First results were already
presented~\cite{Hasenfratz:2005tt,Hashimoto:2006rb,Lang:2006ab,Fukaya:2007fb}. 
The second problem arises in
computations of correlation functions at finite volume in the 
mass range $m\Sigma V\lesssim 1$.
The long-range correlations left in the system generate small 
eigenvalues of the Dirac operator of order $1/V$~\cite{Leutwyler:1992yt}.  
The latter give rise to extremely large fluctuations in 
the commonly used statistical estimators of correlation functions,
and make it difficult to control the statistical 
errors~\cite{Hasenfratz:2002rp,Giusti:2004yp}.

In this paper we address the second problem. We introduce
numerical estimators of the condensate which are infrared 
stable and free from mass-dependent additive ultraviolet 
divergences. This progress builds on the recent 
understanding of the renormalization properties of the spectral 
density of the Dirac operator~\cite{DelDebbio:2005qa}.
The form of these estimators represents and explicit example of how 
the informations from the chiral effective theory can be used not 
only for the interpretation/extrapolation of the data, but also 
to design more efficient numerical algorithms. 

We apply these ideas to the computation of the chiral condensate
in the quenched approximation of QCD with Neuberger fermions.
For cubic lattices with linear extensions $L\gtrsim 1.5$~fm, we
observe a finite-size scaling behaviour of the condensate 
compatible with a theory which exhibits spontaneous 
symmetry breaking at asymptotically large volumes. The topology 
dependence of the data supports the predictions of 
the (quenched) effective theory. Close to the chiral limit
we observe a detailed agreement with the first Leutwyler--Smilga 
sum rule~\cite{Leutwyler:1992yt}. Eventually we extract the value 
of the low-energy constant $\Sigma$ by matching the chiral effective 
theory formulas with the lattice results.

The paper is organized as follows. In Sec.~\ref{sect_cond} we discuss the 
ultraviolet properties of the chiral condensate with Neuberger fermions; 
in Sec.~\ref{sec:chir} we summarize the predictions of the  
effective theory for its volume, mass and topology dependence; 
in Sec.~\ref{sec:lma} we give details of the numerical computation, 
and in Sec.~\ref{sec:phys} we compare our results with the predictions
of the effective theory. The appendices \ref{appa} and \ref{appb} 
are devoted to more technical details.

\section{The quark condensate with Ginsparg--Wilson fermions}\label{sect_cond}
We consider QCD with $N_f\geq 2$ degenerate flavours regularized
on an Euclidean lattice of spacing $a$ with the standard plaquette gauge action 
and with the Neuberger--Dirac operator~\cite{Neuberger:1997fp}. 
The latter (see Appendix A for unexplained notations) satisfies the 
Ginsparg-Wilson relation~\cite{Ginsparg:1981bj}
\begin{equation}
\gamma_5 D+D\gamma_5 =\overline{a}D\gamma_5 D\; ,
\end{equation}
which guarantees an exact chiral symmetry 
\be\label{eq:chisym}
\delta \psi = i \gamma_5\left(1-\overline{a}\,D\right) \psi\; , 
\qquad \delta \overline \psi = \overline \psi \gamma_5 i\; , 
\ee
of the massless fermion action at finite lattice spacing~\cite{Luscher:1998pq}. 
The Jacobian of the 
transformation is non-trivial, and the chiral anomaly 
is recovered {\it \`a la} 
Fujikwa~\cite{Fujikawa:1979ay,Fujikawa:1980eg,Luscher:1998pq}, 
with the topological charge density operator defined 
as~\cite{Neuberger:1997fp,Neuberger:1997bg,Hasenfratz:1998ri}
\be
a^4 q(x) = -\frac{\overline a}{2} \mathrm{tr} \Big[\gamma_5 D(x,x)\Big]\; .
\ee
In a given gauge background the topological charge operator
\be\label{eq:Q}
Q = a^4 \sum_x q(x) 
\ee
is related to the index of the lattice 
Dirac operator $Q = n_+ - n_-$ where $n_+$ ($n_-$) is the number of 
zero modes of $D$ with positive (negative) chirality. All the 
cumulants of the topological charge $Q$ are ultraviolet finite 
and thus the charge distribution~\cite{Giusti:2001xh,Giusti:2004qd,Luscher:2004fu}.
Correlation functions of renormalized local operators 
inserted at a physical distance  are therefore
ultraviolet finite also in the theory restricted to a fixed topological sector. 
The action of the theory is 
invariant under the non-singlet chiral transformations analogous 
to the one in Eq.~(\ref{eq:chisym}). They ensure that in the chiral limit
the quark condensate renormalizes only multiplicatively 
\begin{equation}\label{cond_sim}
\langle \overline{\psi}\psi\rangle = \lim_{a \rightarrow 0}
\; Z_S \langle \overline{\psi}\tilde\psi\rangle\; ,
\end{equation}
where $\tilde{\psi} = \{\tilde\psi_1,\dots,\tilde\psi_{N_f}\}$,
$\overline{\psi}$ is defined analogously and the rotated field 
$\tilde{\psi}_i$ is given in the Appendix A. $Z_S$ is the 
logarithmic-divergent renormalization constant of the scalar
density fixed with a given renormalization prescription.
For our finite-size scaling study we are interested in breaking the 
chiral symmetry explicitly by adding a mass term to the action, which can 
be introduced by defining the massive Dirac operator as 
\begin{equation}
D_m=\left(1-\frac{\overline{a}m}{2}  \right)D+m\; .
\end{equation}
The quark condensate develops additional 
ultraviolet divergences which can be parameterized as 
\be\label{eq:QCDcond}
-\frac{Z_S \langle \overline{\psi}\tilde\psi\rangle}{N_f}  = b_1 m + b_2 m^3 + 
\{\mathrm{finite\;\; terms}\}\; ,
\ee
with the asymptotic behaviour of $b_1$ and $b_2$ being 
$1/a^2$ and $\ln (a)$ respectively.
In finite volume
it is useful to define the chiral condensate in
fixed topological sectors $\nu=|Q|$. It can be parameterized as
\be\label{eq:condnu}
- \frac{\langle \overline{\psi}\tilde\psi\rangle_\nu}{N_f}
= \frac{\nu}{V m} + \chi_\nu\; ,
\ee
where
\be\label{eq:hatchinu}
\hat \chi_\nu = Z_S \chi_\nu = b_1 m + b_2 m^3  + \{\mathrm{finite\;\; terms}\}\; .
\ee
The $1/m$ infrared divergence in Eq.~(\ref{eq:condnu}) is the trivial topology-dependent 
contribution due to the presence of the zero modes of $D$ in the quark propagator. The 
renormalization constant $Z_S$, and the additive subtractions, which are needed to remove
the additive ultraviolet divergences, can be chosen 
to be independent on the topology, see Appendix  \ref{appb}.
As a consequence the continuum limit of the combinations
\be\label{eq:belle}
\hat {\chi}_{\nu_1}- \hat {\chi}_{\nu_2}
\ee 
is unambiguously defined at finite quark mass provided the volume
is large enough.

\indent The spectral density $\rho(\lambda)$ of the massive Dirac 
operator $D_m$ has a well defined thermodynamic limit. It 
can be defined as~\cite{Banks:1979yr} 
\be
\rho(\lambda) = \frac{1}{V}\sum_k \left\langle\, \delta(\lambda-|\lambda_k|)\, \right\rangle
\ee
where $|\lambda_k|^2$ are the non topological eigenvalues of 
$D^\dagger D$ restricted to one of the chiral sectors. Recently it has been shown that 
its renormalized counterpart
\be\label{eq:rhorin}
\hat \rho(\lambda) = Z_S \rho(Z_S \lambda)   
\ee
has a universal continuum limit~\cite{DelDebbio:2005qa}. The same 
applies to the spectral density $\rho_\nu(\lambda)$ defined 
in the theory at fixed topology. At this point the renormalized 
spectral density $\hat \rho(\lambda)$ can, in principle, be computed 
with lattice simulations. However it is more realistic from a numerical
point of view to consider integrals of $\hat \rho(\lambda)$ with
a given probe function. There is clearly a lot of choice. In the rest of our 
paper we will focus on observables ``condensate-like'' of the form
\be\displaystyle\label{eq:chicut}
\hat \tau_\nu(\hat \lambda_\mathrm{min},\hat \lambda_\mathrm{max}) = 
2\hat m {\int_{\hat \lambda_\mathrm{min}}^{\hat \lambda_\mathrm{max}}}
\frac{1-(\bar a Z_S\,\hat \lambda/2)^2}
{[1-(\bar a Z_S\, \hat m/2)^2]\hat\lambda^2 + \hat m^2}\, 
\hat\rho_\nu(\hat \lambda)\, d\hat \lambda\; ,
\ee
with $\hat \lambda = \lambda/Z_S$ and $\hat m = m/Z_S$, which 
have a well defined continuum limit if $\hat \lambda_\mathrm{min}$ 
and $\hat \lambda_\mathrm{max}$ are kept fixed 
when $a\rightarrow 0$. For 
$\hat \lambda_\mathrm{max} \ll \Lambda_{\rm QCD}$ these integrals 
are expected to be dominated from the most infrared part of the 
Dirac spectrum (see below) where also discretization effects are 
expected to be quite small. A more detail study of other 
possible choices is left for future studies. At fixed 
lattice spacing the quantity
\be\label{eq:l2inf}
\hat \tau_\nu(\hat \lambda_\mathrm{min},\infty)  
= \lim_{\hat \lambda_\mathrm{max}\rightarrow \infty} 
\hat \tau_\nu(\hat \lambda_\mathrm{min},\hat \lambda_\mathrm{max})
\ee
has the same ultraviolet divergences of the chiral condensate in 
Eq.~(\ref{eq:hatchinu}). They arise from the contributions 
to the integral at the upper end of the integration domain. 
Since they are topology independent, the observables
\be\label{eq:difftau}
\hat \tau_{\nu_1}(\hat \lambda_\mathrm{min 1},\infty) - 
\hat \tau_{\nu_2}(\hat \lambda_\mathrm{min 2},\infty)
\ee
are unambiguously defined. The chiral condensate 
$\hat \chi_\nu$ is recovered by taking $\hat \lambda_\mathrm{min}=0$
in Eq.~(\ref{eq:l2inf}). The
coefficient of the leading order in $\hat m$ 
is the first Leutwyler--Smilga sum rule~\cite{Leutwyler:1992yt}
\be
\displaystyle
\frac{1}{V}\left. \frac{\partial}{\partial \hat m} \hat {\chi}_{\nu}\right|_{m=0}
= \frac{2}{V} {\int_{0}^{\infty}}\,
\frac{1-(\bar a Z_S\,\hat \lambda/2)^2} 
{\hat\lambda^2}\, \hat\rho_\nu(\hat \lambda)\, d\hat \lambda\; .
\ee
and differences of sum rules analogous to Eq.~(\ref{eq:belle}) are free from
ultraviolet ambiguities. All these considerations remain 
valid in the so-called quenched approximation, where the fermion
determinant is dropped in the effective gluon action.

\section{The quark condensate in the effective chiral theory}\label{sec:chir}
In presence of spontaneous symmetry breaking, the QCD correlation functions 
at small masses and momenta are expected to match those of a chiral effective theory 
which contains only the pseudo-Goldstone bosons as dynamical degrees of 
freedom~\cite{Weinberg:1978kz,Gasser:1983yg}. At asymptotically 
large volumes $L\gg 1/\Lambda_{\rm QCD}$ and for small quark masses
the finite volume behaviour of QCD correlation functions can then be 
predicted~\cite{Gasser:1986vb,Gasser:1987ah}. In the mass
range $\hat m\Sigma V\lesssim \;1$ the partition function of the 
effective theory reads
\be
\mathcal{Z} =\int_{SU(N_f)}dU_0\, {\rm exp}
\left[\mu\,{\rm Re Tr}\,U_0\right]\; ,
\ee
where $U_0$ are the zero-momentum modes in the chiral theory, 
$dU_0$ is the corresponding Haar measure, 
$\mu=\hat m\Sigma V$, and $\Sigma$ is the infinite volume chiral 
condensate. The quark condensate (with opposite sign) is given 
by~\cite{Leutwyler:1992yt}
\be\label{eq:condChPT}
\Sigma(\mu) 
= \frac{\Sigma}{N_f}\frac{\partial}{\partial \mu}\ln \mathcal{Z}
= \frac{\Sigma}{2 N_f}\mu \cdot
\left \{\begin{array}{ll}
2 & N_f=2\\[0.2cm]
1 & N_f \geq 3
\end{array}\right. + \cdots\; ,
\ee
and it vanishes in the chiral limit since spontaneous symmetries breaking 
can only occur in infinite volume. For $\mu\ll 1$ the linear proportionality 
of the condensate to the quark mass and the volume is a distinctive sign of a 
theory which undergoes spontaneous symmetry breaking in infinite 
volume~\cite{Zinn-Justin:2002ru,Banks:1979yr,Leutwyler:1992yt}. It is 
due to the long-range correlations left in the system at finite volume. 
The quark condensate plays the r\^ole of the magnetization of a ferromagnet 
with the quark mass being the analogous of the external magnetic field. 
A tiny variation of the quark mass requires a variation of order $(\hat m V)$ in the
condensate. This behaviour is not spoiled in QCD at finite lattice spacing 
if fermions preserve an exact chiral symmetry. The mass-dependent ultraviolet 
divergences in Eq.~(\ref{eq:QCDcond}) affect only the sub-leading corrections 
suppressed by a factor $1/V$. 

The partition function of the theory at fixed topological 
charge $\nu$ at asymptotically large volumes and in the same mass 
range reads
\be\label{eq:ChPT}
\mathcal{Z}_{\nu}=\int_{U(N_f)}dU\;({\rm det} U)^{\nu}{\rm exp}
\left[\mu\,{\rm Re Tr}\,U\right]\; ,
\ee
and the quark condensate is given by~\cite{Leutwyler:1992yt}
\be\label{eq:Chicond1}
\Sigma_{\nu}(\mu) = \frac{\Sigma \nu}{\mu} + \tilde\chi_\nu(\mu)
 = \frac{\Sigma}{N_f}\frac{\partial}{\partial \mu}\ln \mathcal{Z}_{\nu} = 
 \frac{\Sigma \nu}{\mu} + \frac{\Sigma}{2(N_f + \nu)}\mu + \cdots  \qquad N_f \geq 2
\ee
The infrared divergence matches exactly the analogous one on the 
QCD side in Eq.~(\ref{eq:condnu}). For $\mu\ll 1$ 
the linear proportionality of the condensate to the quark mass and 
the volume signals in this case the presence of a non-zero 
condensate in the thermodynamic limit of the full theory. 
The low-energy constant $\Sigma$ can be determined
with arbitrary precision, if the quark mass and 
the volume $V$ are properly tuned, by matching the combinations 
\be
\tilde \chi_{\nu_1}(\mu) - \tilde \chi_{\nu_2}(\mu)
\ee
with the QCD counterparts in Eq.~(\ref{eq:belle}).
The next-to-leading (NLO) corrections to the scaling behaviour
can be computed in so-called $\epsilon$-expansion~\cite{Gasser:1987ah}. 
They are suppressed 
as $(4\pi F L)^{-2}$, with $F$ being the Goldstone boson decay 
constant in the chiral limit, and they can be included by replacing 
$\Sigma$ with the volume dependent parameter 
$\Sigma_\mathrm{eff}(V)$, whose expression can be found in 
Ref.~\cite{Gasser:1987ah}. 
This implies that the ratios 
\be\label{eq:ratios}
\left. \frac{\tilde{\chi}_{\nu_1}(\mu) - \tilde{\chi}_{\nu_2}(\mu)}
     {\tilde{\chi}_{\nu_3}(\mu) - \tilde{\chi}_{\nu_4}(\mu)} \right|_{\mu=0} 
= \frac{(\nu_1 - \nu_2)}{(\nu_3 - \nu_4)} 
\frac{(\nu_3 + N_f)(\nu_4 + N_f)}
     {(\nu_1 + N_f)(\nu_2 + N_f)}\;, \qquad \nu_3\neq\nu_4\; ,
\ee
are non-trivial parameter free predictions of the effective theory up to NLO, 
which is the higher order at which sub-leading corrections are presently known.
The functional form of the condensate 
$\Sigma_{\nu}(\mu)$ is known for arbitrary values of 
$\mu\leq 1$~\cite{Leutwyler:1992yt,Brower:1980rp}. The condensate 
is also known 
in a partially quenched chiral effective theory with $N_f$ flavors 
and an extra valence one of mass $\hat m_v$. 
From its discontinuity across the imaginary axis
\be\label{eq:disc}
\mathrm{Disc}\Big|_{\hat m_v=i \hat \lambda}\Sigma^\mathrm{pQ}_\nu(\mu,\hat m_v) 
= \lim_{\epsilon \rightarrow 0}
\Sigma^\mathrm{pQ}_\nu(\mu,i\hat \lambda + \epsilon) - 
\Sigma^\mathrm{pQ}_\nu(\mu,i\hat \lambda  + \epsilon)
= 2\pi \tilde \rho_\nu(\hat \lambda) \; ,
\ee
it is possible to extract the NLO effective theory prediction $\tilde \rho(\hat \lambda)$ 
of the spectral density in presence of $N_f$ flavors~\cite{Osborn:1998qb,Damgaard:1998xy}. 
This is the functional 
form that is expected to match $\hat \rho(\hat \lambda)$ in 
Eq.~(\ref{eq:rhorin}) for $\hat \lambda \ll \Lambda_{\rm QCD}\ll 1/\bar a$.
In the following 
for brevity we report the full expressions for the chiral condensate 
and the spectral density for the quenched case only, which is the one 
needed in rest of the paper.

\indent In the quenched approximation of QCD, an effective low-energy
chiral theory is formally obtained if an additional expansion 
in $1/N_c$, where $N_c$ is the number of colors, is carried out 
together with the usual one in quark masses and 
momenta~\cite{Bernard:1992mk,Sharpe:1992ft}. Here we adopt the pragmatic 
assumption that the quenched chiral theory describes the low-energy regime of quenched 
QCD in certain ranges of kinematic scales at fixed $N_c$. 
Correlation functions can be parameterized in terms of effective coupling constants,            
the latter being defined as the couplings of the effective theory. 
\begin{figure}[t]
\begin{center}
\includegraphics[width=7.2cm]{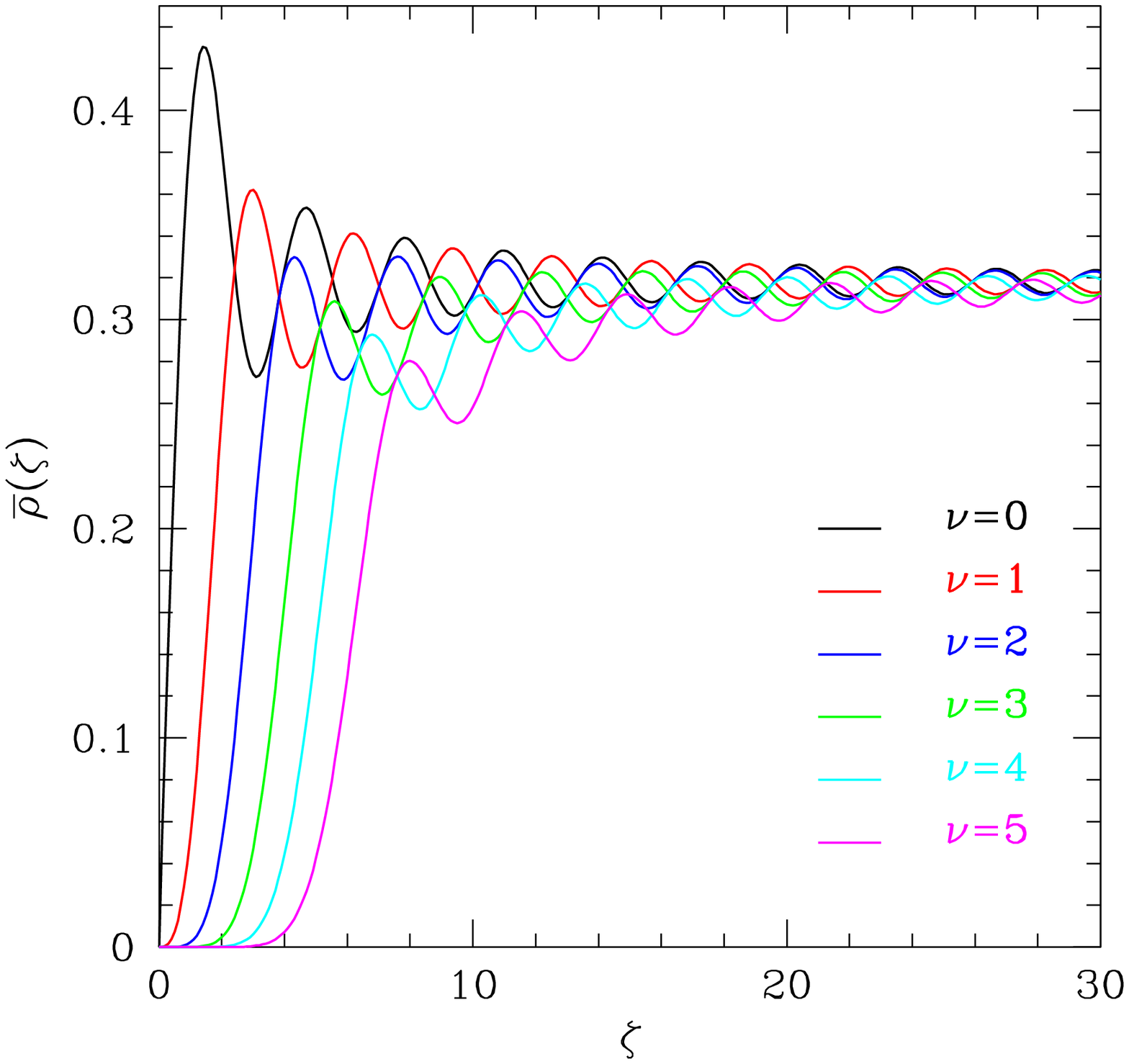}
\hspace{-0.25cm}\includegraphics[width=7.2cm]{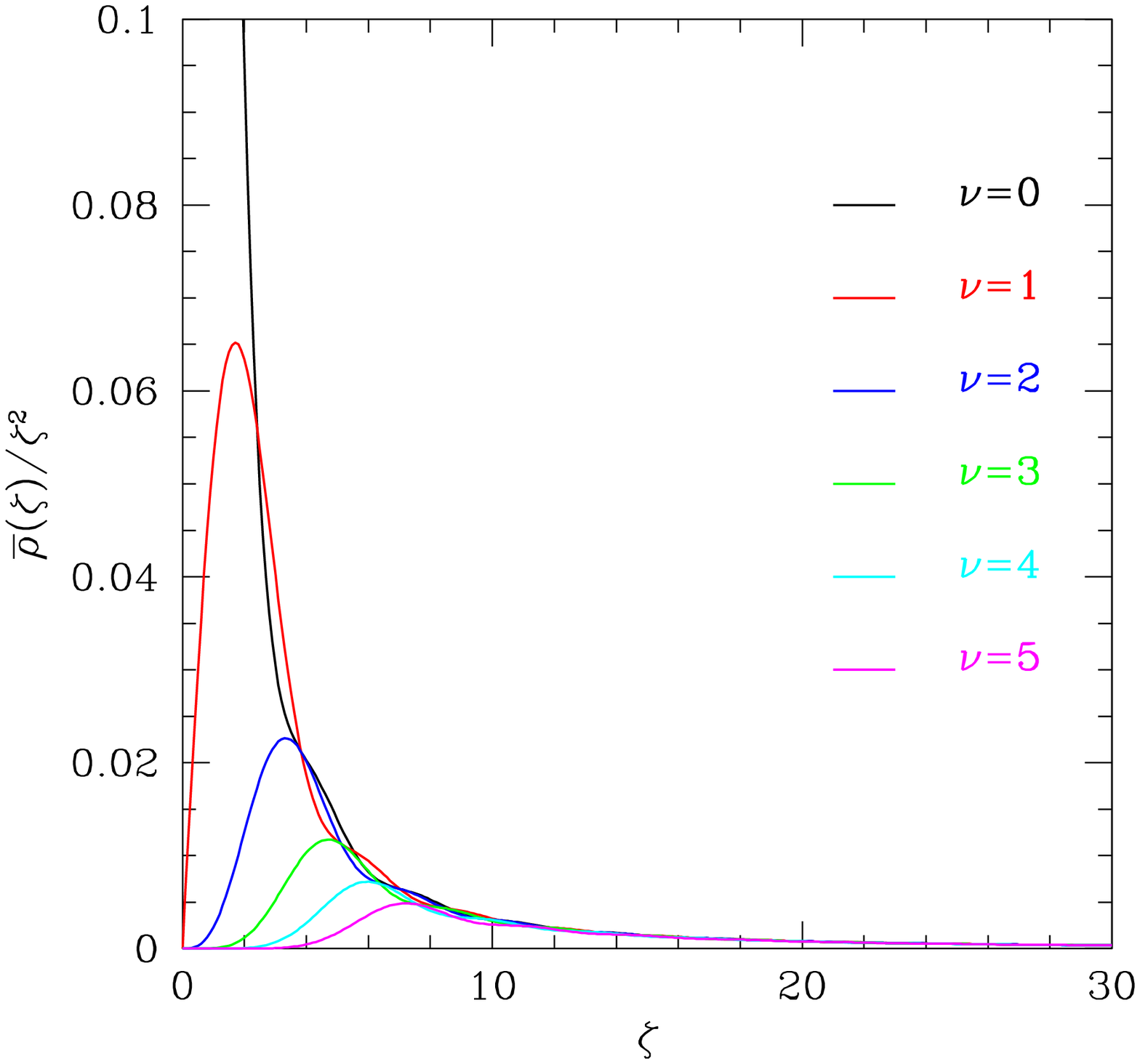}
\caption{The microscopic spectral density and $\bar\rho_\nu(\zeta)/\zeta^2$ for $\nu=0$--$5$ 
are shown on the left and on the right plots respectively.}\label{fig:spectra}
\end{center}
\end{figure}
The quenched condensate at fixed topology $\nu$ at asymptotically 
large volumes (LO) is given by~\cite{Damgaard:1998xy,Osborn:1998qb}
\begin{equation}\label{eq:epsilon_l}
\frac{\Sigma^\mathrm{LO}_{\nu}(\mu)}{\Sigma}=\mu
\left(I_{\nu}(\mu)K_{\nu}(\mu)+ I_{\nu+1}(\mu)K_{\nu-1}(\mu)  \right)+\frac{\nu}{\mu},
\end{equation}
where $I_{\nu}$, $K_{\nu}$ are modified Bessel functions. At the NLO in 
the $\epsilon$-expansion the condensate is~\cite{Osborn:1998qb, Damgaard:2001js}
\begin{equation}\label{eq:epsilon_1l}
\mu\, \Sigma_{\nu}^\mathrm{NLO}(\mu) 
=  \mu_{\rm eff} \Sigma^\mathrm{LO}_{\nu}(\mu_{\rm eff}),
\end{equation}
with $\mu_{\rm eff}=m\Sigma_{\rm eff}V$. The infrared sickness of quenched chiral perturbation theory,
as well as of quenched QCD, manifests itself in the fact that $\Sigma_{\rm eff}$ 
diverges logarithmically with $L$. In the following we define operatively 
$\Sigma_{\rm eff}$ at a given volume as 
\be
\left. \frac{2\nu \tilde \chi_{\nu}(\mu)}{\hat m V}\right|_{\hat m=0} = \Sigma^2_{\rm eff}(V)
\qquad \nu > 0\; . 
\ee
If we consider two box sizes $V_{1,2}=L_{1,2}^4$, the volume dependence of $\Sigma_{\rm eff}$ 
at NLO in $\epsilon$-expansion of the quenched chiral 
theory is~\cite{Damgaard:2001xr,Damgaard:2001js} 
\be\label{eq:fvcor}
\frac{\Sigma_{\rm eff}(V_1)}{\Sigma_{\rm eff}(V_2)} = 1 + \frac{1}{3 F^2}
\left\{ \frac{m_0^2}{8 \pi^2} \ln\left(\frac{L_1}{L_2}\right) - \beta_1 \alpha 
\left(\frac{1}{L_1^2} - \frac{1}{L_2^2}\right)\right\}
\ee
where $\beta_1=0.140461$ is a \emph{shape coefficient}~\cite{Hasenfratz:1989pk}, 
$m_0$ and $\alpha$ are the singlet mass and the additional 
low-energy constant of the singlet kinetic term 
in the chiral Lagrangian respectively (see Ref.~\cite{Damgaard:2001js} 
for unexplained notation). 
From the discussion above it follows that the ratios in Eq.~(\ref{eq:ratios}),
for $\nu_i > 0$ and with $N_f=0$, are non-trivial parameter-free predictions 
also in the quenched effective theory up to the sub-leading 
corrections of higher order presently known. The spectral density of the 
Dirac operator in the quenched approximation can be extracted from Eqs.~(\ref{eq:disc}), 
(\ref{eq:epsilon_l}) and (\ref{eq:epsilon_1l}). They lead 
to the microscopic spectral density~\cite{Osborn:1998qb, Damgaard:2001js}
\be
\bar \rho_\nu(\zeta) =  \frac{\zeta}{2}\left[
J_\nu(\zeta)^2 - J_{\nu+1}(\zeta)J_{\nu-1}(\zeta) \right]\; ,
\ee
which is related to $\tilde \rho_\nu$ as  
\be
\bar \rho_\nu(\zeta) = \frac{1}{\Sigma} 
\tilde \rho_\nu(\zeta/\Sigma V)\; .
\ee
with $\zeta=\hat \lambda \Sigma V$. For $\hat \lambda\ll \Lambda_{\rm QCD}$ 
the integral $\hat \tau_\nu(\hat\lambda_\mathrm{min},\hat\lambda_\mathrm{max})$ 
in Eq.~(\ref{eq:chicut}) can then be matched with 
\be\displaystyle\label{eq:tauChPT}
\tilde \tau_\nu\left(\frac{\zeta_\mathrm{min}}{\Sigma V},\frac{\zeta_\mathrm{max}}{\Sigma V}\right) = 
2\Sigma\mu {\int_{\zeta_\mathrm{min}}^{\zeta_\mathrm{max}}}
\frac{1}{\zeta^2 + \mu ^2}\, 
\bar\rho_\nu(\zeta)\, d\zeta\; ,
\ee
The functions $\bar\rho_\nu(\zeta)$ and $\bar\rho_\nu(\zeta)/\zeta^2$ are 
shown in Fig.~\ref{fig:spectra}. They highlights the fact that 
large parts of the integrals in Eq.~(\ref{eq:tauChPT})
come from the region with $\zeta\lesssim 20$, where a handful of low-lying 
eigenvalues of the Dirac operator provides the bulk of the contribution.
Similar conclusions apply in the range $\mu\leq 1$.

\section{Numerical computation}\label{sec:lma}
The primary observables in our numerical computations 
are the integrals $\hat \tau_\nu(\hat \lambda_\mathrm{min},\infty)$ 
defined in Eq.~(\ref{eq:l2inf}). The lower 
bound $\hat \lambda_\mathrm{min}$ is chosen 
to ensure the reliability of the Monte Carlo 
estimate (see below) and at the same time 
to minimize the difference with $\hat {\chi}_{\nu}$. 
The low energy constant $\Sigma_\mathrm{eff}$ is extracted  
from a matching of the combinations in Eq.~(\ref{eq:difftau}) 
with the corresponding formulas in the effective theory. 
Our final goal, the computation of  
$(\hat {\chi}_{\nu_1}- \hat {\chi}_{\nu_2})$,
is then achieved by adding the small contribution
from the infrared tail analytically using the 
functional form in Eq.~(\ref{eq:tauChPT}).

\subsection{Numerical estimator}
The underlying chiral symmetry guarantees that
for every background gauge configuration the quark 
propagator $S_m(x,y)$ (see Appendix~\ref{appa}) 
satisfies
\be
\sum_x {\rm tr}\, \Big[S_m(x,x)\Big] = 
\frac{\nu}{m} + 2 \sum_x {\rm tr} \Big[P_c\,  S_m(x,x)\, P_c\Big]
\ee
with $\nu$ being the absolute value of the topological charge of the 
configuration and 
$P_c$ the chiral projector into the chiral sector without zero 
modes\footnote{In finite volume the probability of having 
a configuration with zero modes in both chiralities is zero.}. 
In this form, and once the topological charge is known, the calculation of 
the quark condensate requires the computation of  
the propagator in the sector without zero modes only~\cite{Hernandez:1999cu}.

\noindent On reasonably large lattices the computation of  
all eigenvalues of the Dirac operator or of the propagator 
from every source point $x$ is 
numerically unfeasible. It is quite standard, however, 
to compute the quark propagator from a few fixed source points or to extract a 
few low-lying eigenvectors. Let us assume 
that once $\nu$ has been determined, a number $n$ of approximate 
low-lying eigenvalues of $P_c D^\dagger DP_c$ and the corresponding eigenvectors is 
computed by minimizing the Ritz functional starting from random vectors 
generated with a Gaussian action, i.e. invariant under 
space-time translations~\cite{Giusti:2002sm}. 
The minimization is carried out until the approximated eigenvalues 
have a relative error $\omega_k$ and satisfy
\begin{eqnarray}\label{eq:ritz}
P_c D^\dagger D P_c u_k & = &|\overline \lambda_k|^2 u_k+r_k\; ,\;\;\;(k=1,...,n)\; ;\\
(u_l,u_k) &= & \delta_{lk}\; ;\\
(u_l,r_k) &= & 0\;\; \forall (l,k)\; ,\;\;\;||r_k||\leq \omega_k |\overline \lambda_k|^2\; .
\end{eqnarray}
The propagator in the sector without zero modes can then be split
in a \emph{light} and a \emph{heavy} contribution as follows:
\be\label{eq:split}\displaystyle
P_c\,  S_m(x,y)\, P_c =  m\, \sum_{k=1}^n \frac{P_c \tilde{u}_k(x) u^\dagger_k(y) P_c}{(1-\overline a^2 m^2/4)|\overline \lambda_k|^2 + m^2} + P_c S^h(x,y)P_c \nonumber\; ,
\ee
where $ \tilde{u}_k$ is defined from $u_k$ as in Eq.~(\ref{eq:psitilde}). It is easy
to prove that, once averaged over the gauge configurations, the spin-color trace
of each 
contribution on the right-hand side of Eq.~(\ref{eq:split}) is 
translational invariant even if the $u_k$ are 
only approximate eigenvectors, i.e. $\omega_k\neq 0$~\cite{Giusti:2002sm}. The condensate,
after the trivial contribution from the zero modes is subtracted, can then be decomposed as 
\be\label{eq:condfin}
{\chi}_{\nu}=  \frac{1}{V}   
\sum_{x} \left\langle {\rm tr}\, \Big[S_m(x,x)\Big]\right\rangle_\nu - \frac{\nu}{m V}= \chi_\nu^l+\chi_\nu^h\; ,
\ee
where the heavy and the light contributions can be computed as 
\bea
\chi_\nu^l & = & \frac{2\,m}{V} \sum_{k=1}^n \left\langle 
\frac{1-\overline a^2|\overline \lambda_k|^2/4}{(1-\overline a^2 m^2/4)|\overline \lambda_k|^2 + m^2}\right\rangle_\nu\; ,
\label{eq:condhl}\\[0.25cm]
\chi_\nu^h & = &  2\, \left\langle {\rm tr}\left[P_c S^h(0,0)P_c  \right]\right\rangle_\nu \; .\label{eq:condhh}
\eea
It must be stressed that Eq.~(\ref{eq:condfin}) {\it is exact independently on the values
of $\omega_k$ and the number of the extracted eigenvectors $n$}. By contrast, the statistical 
variance of the estimator changes with $\omega_k$ and $n$. The local fluctuations of the 
approximated eigenvectors $u_k(x)$ are enhanced on the r.h.s. of Eq.~(\ref{eq:split}) 
by the smallness of the denominator. Under the working assumption that they 
are responsible for large variations in the trace of the local propagator, 
the variance of the estimator in Eq.~(\ref{eq:condfin}) is greatly reduced with respect 
to the one of ${\rm tr}\left[P_c S(0,0)P_c  \right]$. Deep in the chiral regime
the probability of having a configuration with eigenvalues lying in the infrared tail of the 
spectral density $\hat \rho(\lambda)$ is small but not negligible. The tail is poorly 
sampled by the Monte Carlo while at the same time the smallness of the eigenvalues generates large
fluctuations in the light part of the condensate $\chi_\nu^l$. To overcame this problem
we replace $\chi_\nu^l$ with 
\be
\tau_\nu^l  = \frac{2\,m}{V} \sum_{k=1}^n \left\langle 
\frac{1-\overline a^2|\overline \lambda_k|^2/4}{(1-\overline a^2 m^2/4)|\overline \lambda_k|^2 + m^2}
\theta(|\lambda_k| - \lambda_\mathrm{min})\right\rangle_\nu\; ,
\ee
in Eq.~(\ref{eq:condfin}), where $\theta$ is the usual 
step function. The value of $\lambda_\mathrm{min}$
is chosen in such a way that only 5 eigenvalues  
in all the Monte Carlo history of a given data set 
are lower. The parameter $\Sigma_\mathrm{eff}$ is then 
extracted by matching these observables with 
the analogous ones in the effective theory. Eventually the chiral 
condensate $\chi_\nu$  is computed by adding the small contribution from the 
tail analytically using the formula in Eq.~(\ref{eq:tauChPT}).
When the latter is substantial the determination 
of $\Sigma_\mathrm{eff}$ from differences of 
$\hat \tau_\nu(\hat \lambda_\mathrm{min},\infty)$ is still correct,
but the value of $\chi_\nu$ is heavily affected by 
the functional form used and therefore 
it less interesting to us. In the following we only consider data sets
where the integral of the tail is at most 25 per cent of 
$\tilde \chi_\nu$, and in most of the cases it is 
less than 10 per cent. A positive side effect of this requirement is 
that the computation of the eigenvalues does not have 
to be very precise. The tail contribution in $\chi_\nu$ can, of course,
be reduced by increasing the statistics of the data set, i.e. sampling
part of the tail with confidence.

This procedure complements for the case of the chiral condensate 
the low-mode averaging (LMA) technique proposed in Ref.~\cite{Giusti:2004an}. 
The latter has been successfully applied already to meson two-point 
functions \cite{Giusti:2004an,Fukaya:2005yg,Ogawa:2005jn}, baryon two-point 
functions \cite{Giusti:2005sx} and more recently to three-point functions for 
the extraction of low-energy constants of the $\Delta S=1$ chiral effective 
Hamiltonian \cite{Giusti:2006mh}.

\subsection{Numerical experience}
%%%%%%%%%%%%%%%%%%%%%%%%%%%%%%%%%%%%%%%%%%%%%%%%%%%%%%%%%%%%%%%%%%%%%%%%%%%%%%%%%%%%%%%%%%%%%%%%%%%%%%%%%%%%%%%%%%%
\begin{table}[t]
\begin{center}
\begin{tabular}{|c| c c c c|c|c|}
\hline
lat & $\beta$  &  $L/a$ & $L$ (fm)  &  $N_\mathrm{cfg}$ & $N^\nu_\mathrm{cfg}$  & $am$  \\
\hline
c1 & 5.8458   & 12 & 1.49 fm & 672       &  119, 205, 155, &  0.001, 0.003, 0.008,\\
    &          &   &     &           &     104, 51, 29&  0.012 ,0.0016\\ 
 \hline
c2 & 5.8458   & 16 &  1.98 fm     & 488       &  49, 69, 82, &  0.000316, 0.000949, 0.00253, \\
    &          &   &     &           & 72, 50, 54 &  0.00380, 0.00506 \\
\hline
c3 & 6.0      & 16 &  1.49 fm  & 418     & 74, 137, 101, &  0.000612, 0.00184, 0.00490, \\
    &          &   &     &           &  62, 27, 12 &  0.00735, 0.00980 \\
\hline
\end{tabular}
\caption{Parameters of the simulations: $\beta=6/g^2$ is the bare gauge coupling, 
$L$ is the linear extent of the each lattice, $N_\mathrm{cfg}$ is the total number of 
configurations generated, and $am$ are the bare quark masses considered. $N^\nu_\mathrm{cfg}$ 
refers to the subset of configurations with fixed topological charge for $\nu=0$--$5$.
}\label{tab:simul}
\end{center}
\end{table}
%%%%%%%%%%%%%%%%%%%%%%%%%%%%%%%%%%%%%%%%%%%%%%%%%%%%%%%%%%%%%%%%%%%%%%%%%%%%%%%%%%%%%%%%%%%%%%%%%%%%%%%%%%%%
%%%%%%%%%%%%%%%%%%%%%%%%%%%%%%%%%%%%%%%%%%%%%%%%%%%%%%%%%%%%%%%%%%%%%%%%%%%%%%%%%%%%%%%%%%%%%%%%%%%%%%%%%%%%%%%%%%%%%%%%
\begin{figure}[t]
\hspace{-0.25cm}\includegraphics[width=7.5cm]{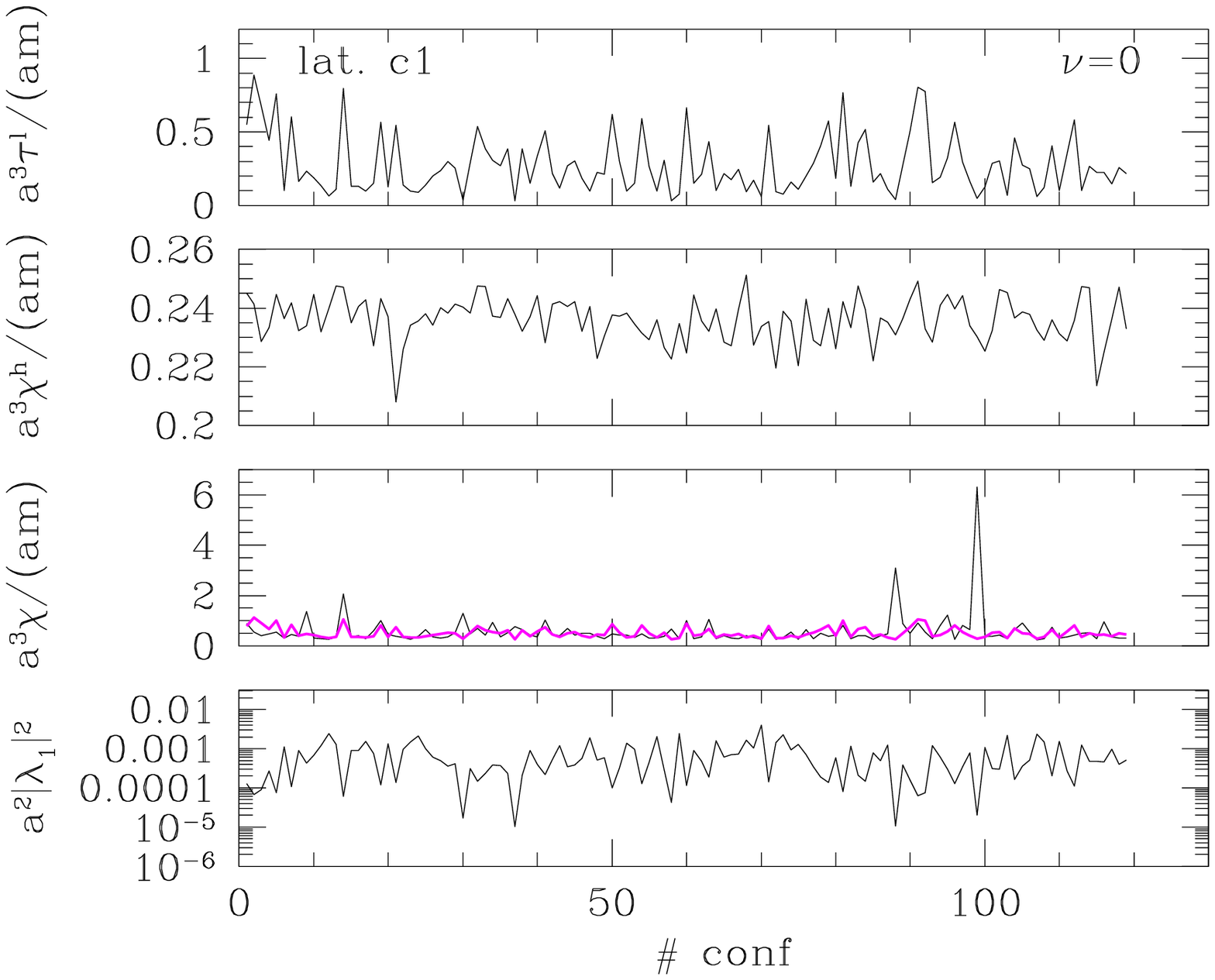}
\hspace{-0.25cm}\includegraphics[width=7.5cm]{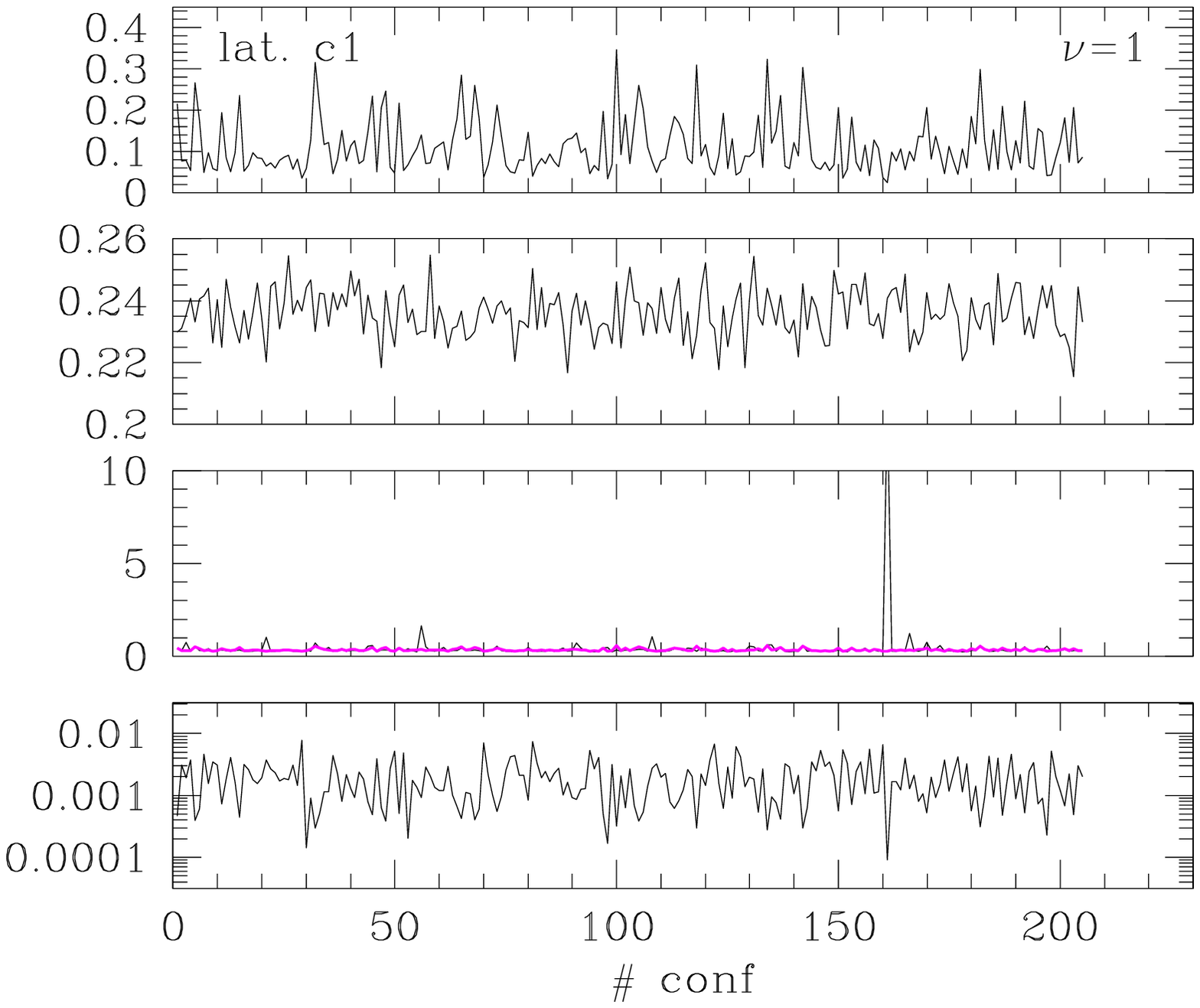}
\caption{Monte Carlo history for the run c1, $am=0.008$ and for the topological 
sectors $\nu=0$ (left) and  $\nu=1$ (right). 
The first plot from the bottom shows the Monte Carlo history of the absolute value of the 
smallest eigenvalue of $P_cD^\dagger D P_c$; the second one represents the condensate 
(divided by the quark mass) computed without LMA (black-thin line) and 
the sum $a^3(\tau_\nu^l + \chi_\nu^h)/(am)$ (magenta-thick line). 
The third and the fourth show the Monte Carlo histories of 
the heavy contribution $(a^3\chi_\nu^h)/am$ and the light one 
$(a^3\tau_\nu^l)/am$, respectively. }\label{hist_c1}
\end{figure}
%%%%%%%%%%%%%%%%%%%%%%%%%%%%%%%%%%%%%%%%%%%%%%%%%%%%%%%%%%%%%%%%%%%%%%%%%%%%%%%%%%%%%%%%%%%%%%%%%%%%%%%%%%%%%%%%%%%%%%%%%
We have generated ensembles of gauge configurations
by standard Monte Carlo techniques with the Wilson gluon action 
and periodic boundary conditions. The fermions are discretized with 
the Neuberger--Dirac operator as defined in Eq.~(\ref{eq:neu}) with $s=0.4$.
A summary of the parameters of our runs are reported in Tab.~\ref{tab:simul}. 
We have simulated three lattices with two different volumes (runs c1 and c2) and two 
lattice spacings (runs c1 and c3). The linear extent always satisfies 
$L \gtrsim 1.5$~fm, a size that we expect to be large enough 
for a finite-size scaling study. This is suggested by the results in Ref.~\cite{Giusti:2003gf}, 
where for  $L \gtrsim 1.5$~fm it was found a detailed agreement of the predictions of random matrix theory 
for the low-lying eigenvalues of the Dirac operator with quenched QCD results. The values of the two lattice 
spacings are chosen to guarantee the locality of the Neuberger operator~\cite{Hernandez:1998et}, and to be 
in a range where discretization effects were found to be small in several 
observables~\cite{Babich:2005ay,Wennekers:2005wa}. The quark masses for the lattice c1 are fixed to be roughly 
in the interval $0.07\;\lesssim \;\mu\;\lesssim \;1.2$. For the lattice c2 the masses are such that
the values of $(mV)$ match those of c1. The masses for the lattice c3 are chosen so that 
the dimensionless quantity $(mV/\hat Z_S r_0^3)$ is constant, where
$r_0=0.5$~fm is a widely used reference scale in quenched QCD 
computations~\cite{Sommer:1993ce,Guagnelli:1998ud} and $\hat Z_S$ 
is the renormalization constant of the scalar density in the 
RGI scheme which we have taken from Ref.~\cite{Wennekers:2005wa}. 
The calculation of the topological charge, of the low-lying eigenvalues, and 
of the quark propagator is performed following Ref.~\cite{Giusti:2002sm}. For each run 
the low-mode averaging is implemented as described in the previous subsection with
$n=20$ and $\omega_k=0.05$. A posteriori we have verified that the highest eigenvalue
extracted satisfies $(\langle |\overline \lambda_{20}|\rangle \Sigma_\mathrm{eff} V) > 20$ for all lattices.

In Fig.~\ref{hist_c1} we show a typical Monte Carlo history 
for the smallest eigenvalue of the Neuberger operator, for the chiral condensate
with and without LMA and for the heavy and light contributions
separately. It is obtained from the run c1 for $\nu=0,1$ and $am=0.008$. A first observation is that the heavy part of 
the condensate is very stable in all topological sectors. Moreover the quantity 
${\rm tr}\left[P_c S^h(0,0)P_c  \right]/(am)$ is essentially independent on the quark mass 
{\it configuration by configuration}. For all lattices, its largest relative deviation 
that we have observed among different masses is roughly $10^{-3}$. We interpret this as 
a consequence of the fact that $|\overline \lambda_{20}|\gg m$ configuration by configuration. 
We thus expect that a further stabilization of the heavy contribution would not reduce the variance 
of the condensate significantly. The Monte Carlo history of $\tau_\nu^l$ does not show 
large spikes for all masses, and a statistical analysis is applicable. Some of the lighter 
masses at lower
topologies, however, have been discarded (see Tab.~\ref{tab:res}) to satisfy the upper limit on the tail contribution
discussed in the previous subsection. As expected the light part fluctuates much more than the heavy one.
Large contributions 
appear in coincidence with the lower values of $|\lambda_1^2|$ consistently with the expectations. 
The Monte Carlo history of the local estimator ${\rm tr}\left[P_c S(0,0)P_c  \right]$
has fluctuations which are much larger than those with LMA. Moreover for the lower topologies
extreme statistical fluctuations are observed for most of the masses considered, 
which invalidate the statistical analysis of the sample. In the following the physics analysis 
is carried out only on the data with LMA.
%%%%%%%%%%%%%%%%%%%%%%%%%%%%%%%%%%%%%%%%%%%%%%%%%%%%%%%%%%%%%%%%%%%%%%%%%%%%%%%%%%%%%%%%%%%%%%%%%%%%%%%%%%%%%%%%%%%%%%%
\begin{figure}[!ht]
\begin{center}
\includegraphics[width=10.0cm]{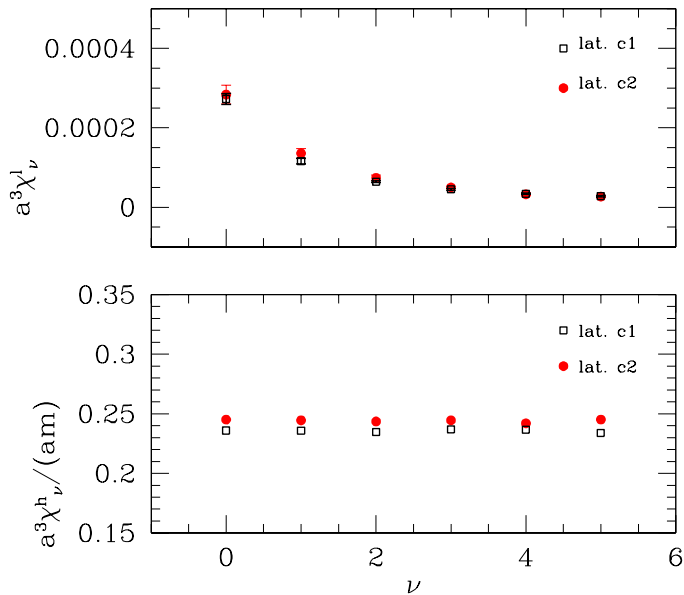}
\caption{The quantities $a^2 \chi^h_{\nu}/m$ (bottom) and $a^3\chi^l_{\nu}$ (top) as a function of 
$\nu$, for the second heaviest mass of the run c1 (black points) and run c2 (red points).}\label{fig:cond_h}
\end{center}
\end{figure}
%%%%%%%%%%%%%%%%%%%%%%%%%%%%%%%%%%%%%%%%%%%%%%%%%%%%%%%%%%%%%%%%%%%%%%%%%%%%%%%%%%%%%%%%%%%%%%%%%%%%%%%%%%%%%%%%%%%%%%%
%%%%%%%%%%%%%%%%%%%%%%%%%%%%%%%%%%%%%%%%%%%%%%%%%%%%%%%%%%%%%%%%%%%%%%%%%%%%%%%%%%%%%%%%%%%%%%%%%%%%%%%%%%%%%%%
\begin{figure}[!ht]
\includegraphics[width=7.5cm]{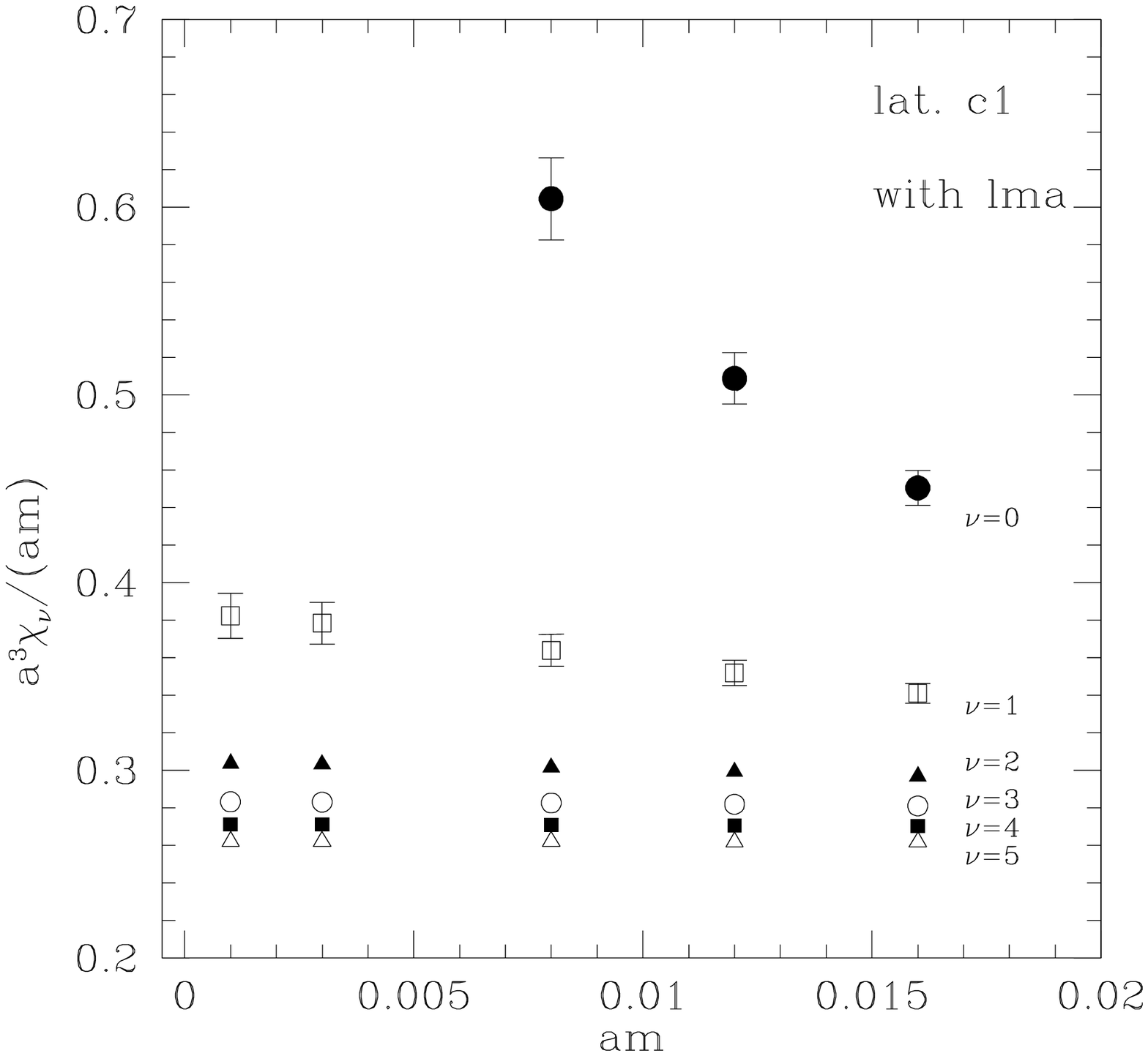}
\hspace{-0.375cm}\includegraphics[width=7.5cm]{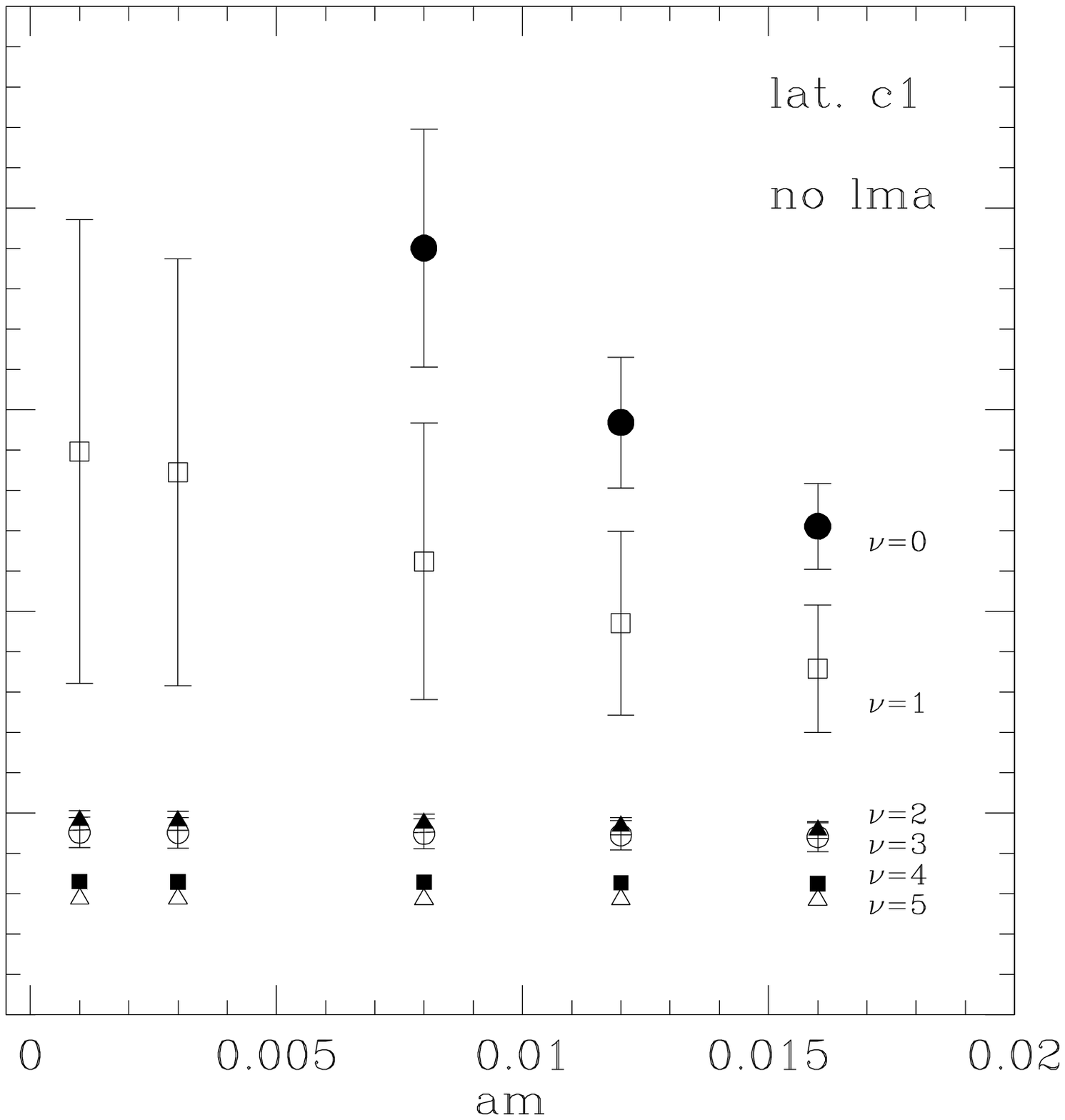}
\caption{Results for $(a^3\chi_\nu)/am$ as a function of the quark mass, for the lattice c1 at several values of $\nu$.
The data on the left are obtained with low-mode averaging, while on the right we show the values computed with the local estimator.}\label{fig_cond}
\end{figure}
%%%%%%%%%%%%%%%%%%%%%%%%%%%%%%%%%%%%%%%%%%%%%%%%%%%%%%%%%%%%%%%%%%%%%%%%%%%%%%%%%%%%%%%%%%%%%%%%%%%%%%%%%%%%%%%%%%%%%%%%%%%

In Fig.~\ref{fig:cond_h} we show 
our results for the heavy and the light contributions to the chiral condensate as 
a function of $\nu$ for the second heaviest mass of the runs c1 and c2. The heavy part divided 
by the quark mass is weakly dependent on the topological charge and the volume.
The results for $a^3\chi^l_{\nu}$ at the corresponding masses are compatible. 
This indicates that the light contribution is to a good approximation a function 
of the variable $(mV)$. In addition, the $1/\nu$ behavior predicted  close to the chiral limit 
by the first Leutwyler--Smilga sum rule is reproduced qualitatively. 
These observations, which are valid for all masses simulated, point to the fact that 
the splitting between heavy and light contribution is such that the bulk of the heavy part behaves 
essentially like 
an ultraviolet divergence (i.e. volume and topology independent and linear in the mass), 
while the light part scales with respect to topology, mass and volume essentially as predicted 
by chiral perturbation theory. The analysis in the chiral effective theory in the previous
section suggests that these features are mostly volume independent.

The effect of low-mode averaging can be appreciated 
in Fig. \ref{fig_cond}, where we show $(a^3\chi_\nu)/am$
for $\nu=0$--$5$ as a function of the quark mass, with (left) and without (right) 
LMA\footnote{At lower topologies and without LMA the average values 
and their errors are only indicative since the statistical analysis is 
invalidated by extreme statistical fluctuations.}. It is clear that 
the variance reduction is much more effective for the sectors with lower topological charge, which are 
dominated by infrared contributions. Nevertheless with LMA we still obtain a variance reduction of a 
factor $\sim 2$ up to $\nu=5$ in all our runs.

%%%%%%%%%%%%%%%%%%%%%%%%%%%%%%%%%%%%%%%%%%%%%%%%%%%%%%%%%%%%%%%%%%%%%%%%%%%%%%
\begin{table}[t]
\begin{center}
\setlength{\tabcolsep}{.2pc}
\begin{tabular}{|c |l| c c c c c|}
\hline
lat & $\nu \backslash am$ & 0.001 & 0.003 & 0.008 & 0.012 & 0.016\\
\hline
c1  &   0    &               &              & 0.00484(18)   & 0.00610(16) & 0.00721(15)\\
    &   1    & 0.000382(12)  & 0.00114(3)   & 0.00291(7)    & 0.00422(8)  & 0.00546(8)  \\
    &   2    & 0.000304(3)   & 0.000910(8)  & 0.002413(19)  & 0.00359(3)  & 0.00475(3) \\     
    &   3    & 0.000283(3)   & 0.000849(8)  & 0.002261(21)  & 0.00338(3)& 0.00450(4) \\
    &   4    & 0.0002712(22) & 0.000814(7)  & 0.002168(17)  & 0.00325(3)& 0.00433(3)\\
    &   5    & 0.0002622(23)   & 0.000787(7)  & 0.002097(18)  & 0.00314(3)  & 0.00419(4)\\
\hline
\hline
lat &  $\nu \backslash am$& 0.000316 &  0.000949   &0.00253 & 0.00380  & 0.00506\\
\hline
c2  &   0    &               &              &              & 0.0043(3)    &  0.00480(25)\\
    &   1    &               &              & 0.00182(12)  & 0.00256(14)  &  0.00319(15) \\
    &   2    & 0.000157(9)   & 0.00047(3)   & 0.00123(6)   &   0.00182(9) &  0.00237(10)\\
    &   3    & 0.000129(3)   & 0.000387(8)  & 0.001026(20) & 0.00153(3)   &  0.00202(4)\\
    &   4    & 0.000110(4)   & 0.000329(11) & 0.00088(3)   & 0.00131(4)   &  0.00174(5)\\
    &   5    & 0.0001047(21) & 0.000314(6)  & 0.000837(16) & 0.001255(24) &  0.00167(3)\\
\hline
\hline
lat & $\nu \backslash am$ & 0.000612 &  0.00184   &0.00490 & 0.00735  & 0.00980\\
\hline
c3  &   0    &             &             & 0.00246(11) & 0.00324(11)  & 0.000394(10)\\
    &   1    & 0.000217(7) &0.000648(19) & 0.00168(4)& 0.00246(5)&0.00321(5) \\
    &   2    &  0.0001850(14)&0.000556(4) &0.001474(11) & 0.002200(15) &0.002915(19) \\
    &   3    & 0.0001729(18)& 0.000520(5)  & 0.001382(14) & 0.002070(20)  & 0.00276(3) \\
    &   4    & 0.0001677(17)&   0.000504(5) & 0.001342(13) & 0.002012(20)&    0.00268(3)\\
    &   5    & 0.0001622(17)& 0.000488(5) & 0.001299(13) & 0.001947(20) & 0.00260(3) \\
\hline
\end{tabular}
\caption{Numerical results for $a^3 \chi_{\nu}$ with $\nu=0$--$5$.}\label{tab:res}
\end{center}
\end{table}
%%%%%%%%%%%%%%%%%%%%%%%%%%%%%%%%%%%%%%%%%%%%%%%%%%%%%%%%%%%%%%%%%%%%%%%%%%%%%%
\section{Comparison with the effective theory}\label{sec:phys}
In this section we compare our numerical results with the predictions of 
the chiral effective theory. We first focus on the finite-size scaling
and the topology dependence of the condensate, and then we extract the value 
of $\Sigma_{\rm eff}$ from a fit in the mass of our data. Our raw numerical results
of $\chi_{\nu}$ for $\nu=0$--$5$, are reported in Table~\ref{tab:res}. 
The statistical errors have been estimated with a jackknife procedure.

\subsection{Finite-size scaling}\label{sec:fss}
\begin{figure}[!ht]
\hspace{-0.25cm}\includegraphics[width=7.5cm]{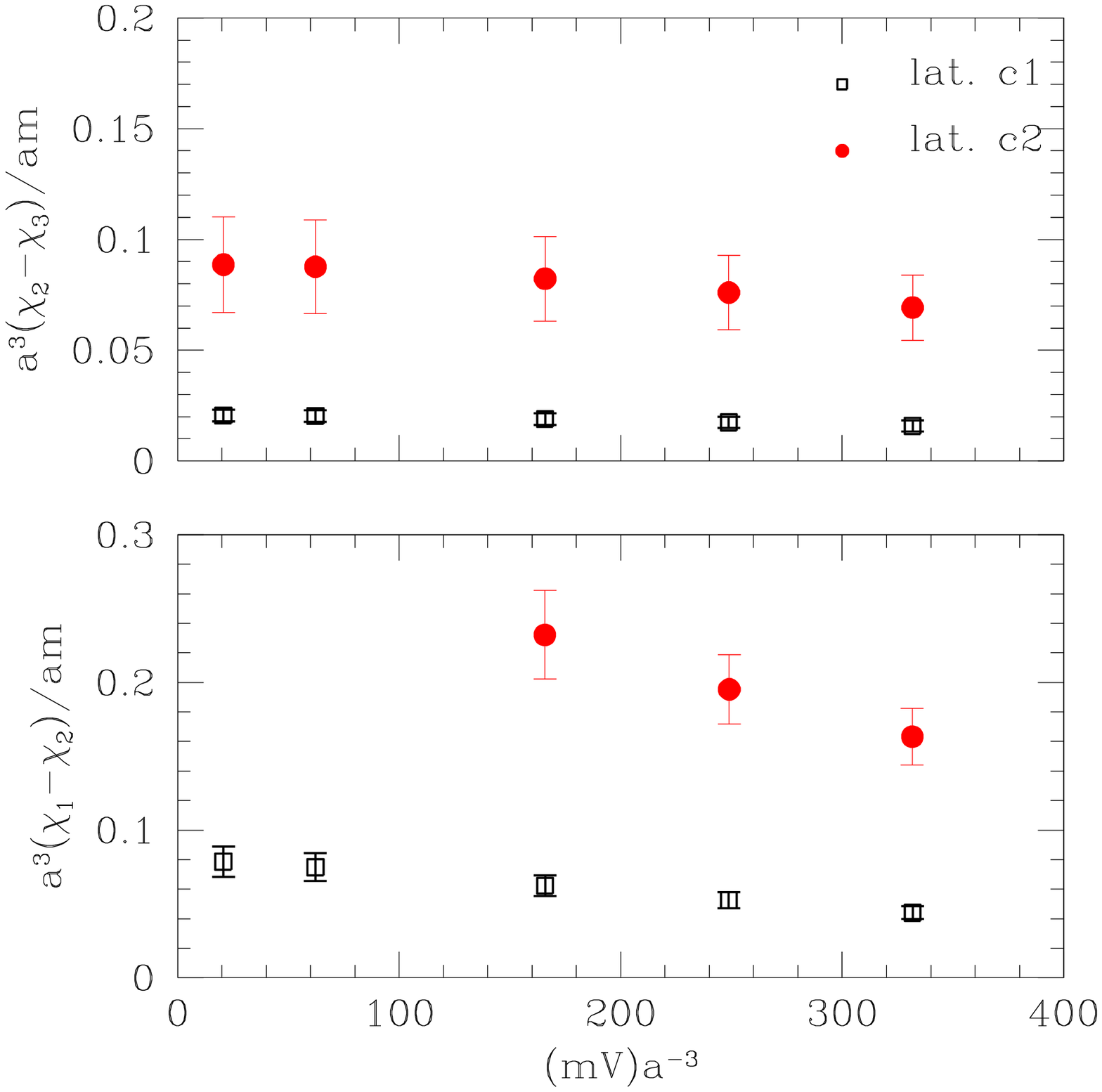}
\hspace{-0.25cm}\includegraphics[width=7.5cm]{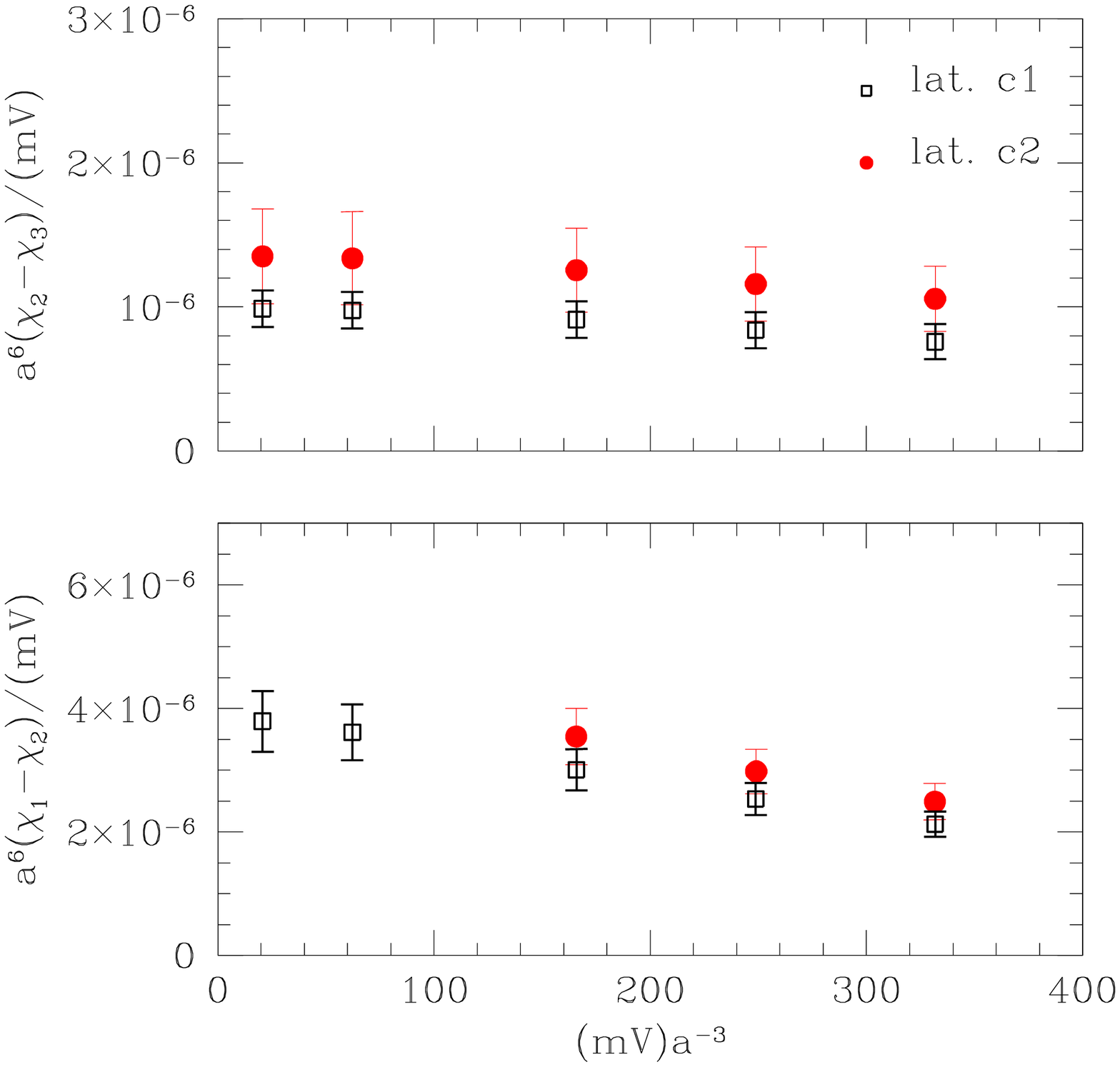}
\caption{The ratios $a^3 (\chi_{1} - \chi_{2})/(am)$  and  $a^3 (\chi_{2} - \chi_{3})/(am)$ 
(left), and $a^6(\chi_{1}-\chi_{2})/(mV)$ and $a^6 (\chi_{2} - \chi_{3})/(mV)$ (right) 
as a function of $(m V)a^{-3}$ for lattices 
c1 and c2.}\label{fig:FSS}
\end{figure}

The parameters of the lattices c1 and c2 have been chosen
to carry out a finite-size scaling study. The 
lattice spacing is the same and the physical 
volume differ by more than a factor $3$.  The bare 
combinations $a^3 (\chi_{1} - \chi_{2})/am$ and 
$a^3 (\chi_{2} - \chi_{3})/am$ computed
on these lattices are shown in the plots on the left 
of Fig.~\ref{fig:FSS}. Their values differ from zero 
by many standard deviations for both volumes. Finite 
size effects are clearly visible. The rescaled combinations
are shown in the plots on the right of the same Figure as a 
function of $(m V)a^{-3}$. The  corresponding
data sets are in very good agreement for all masses simulated
within the statistical errors. These results are compatible
with the finite size scaling behaviour expected at asymptotically 
large volumes for a theory which exhibits spontaneous 
symmetry breaking. Within our statistical errors we do not observe 
deviations from the leading scaling behaviour. Similar conclusions 
apply for the other combinations made with $\nu=0$--$3$. Results
from higher topological sectors tend to depart from the LO scaling behaviour, 
but our statistical errors are too large to draw any definite conclusion. 
\begin{figure}[ht]
\hspace{-0.25cm}\includegraphics[width=7.5cm]{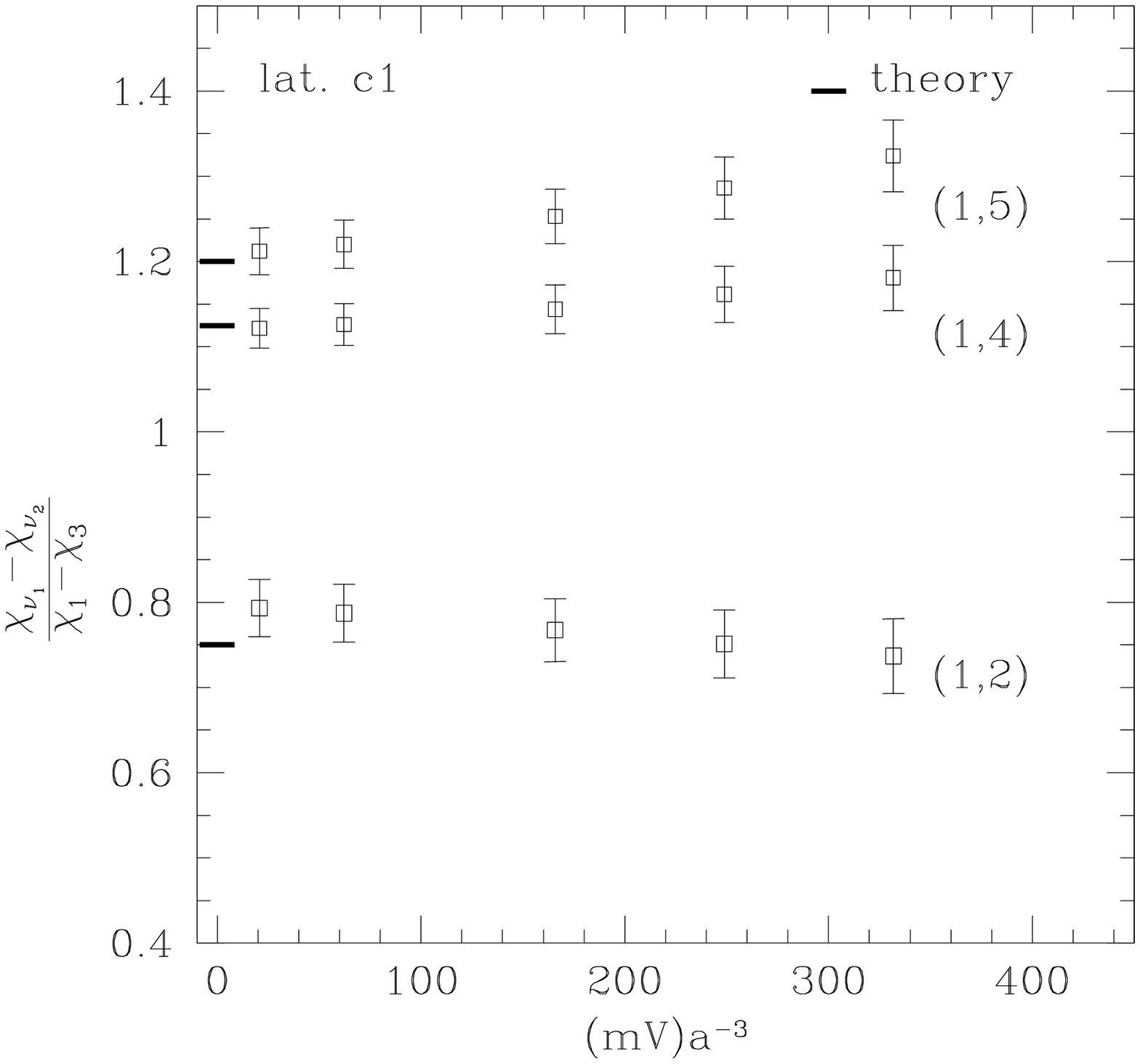}
\hspace{-0.25cm}\includegraphics[width=7.5cm]{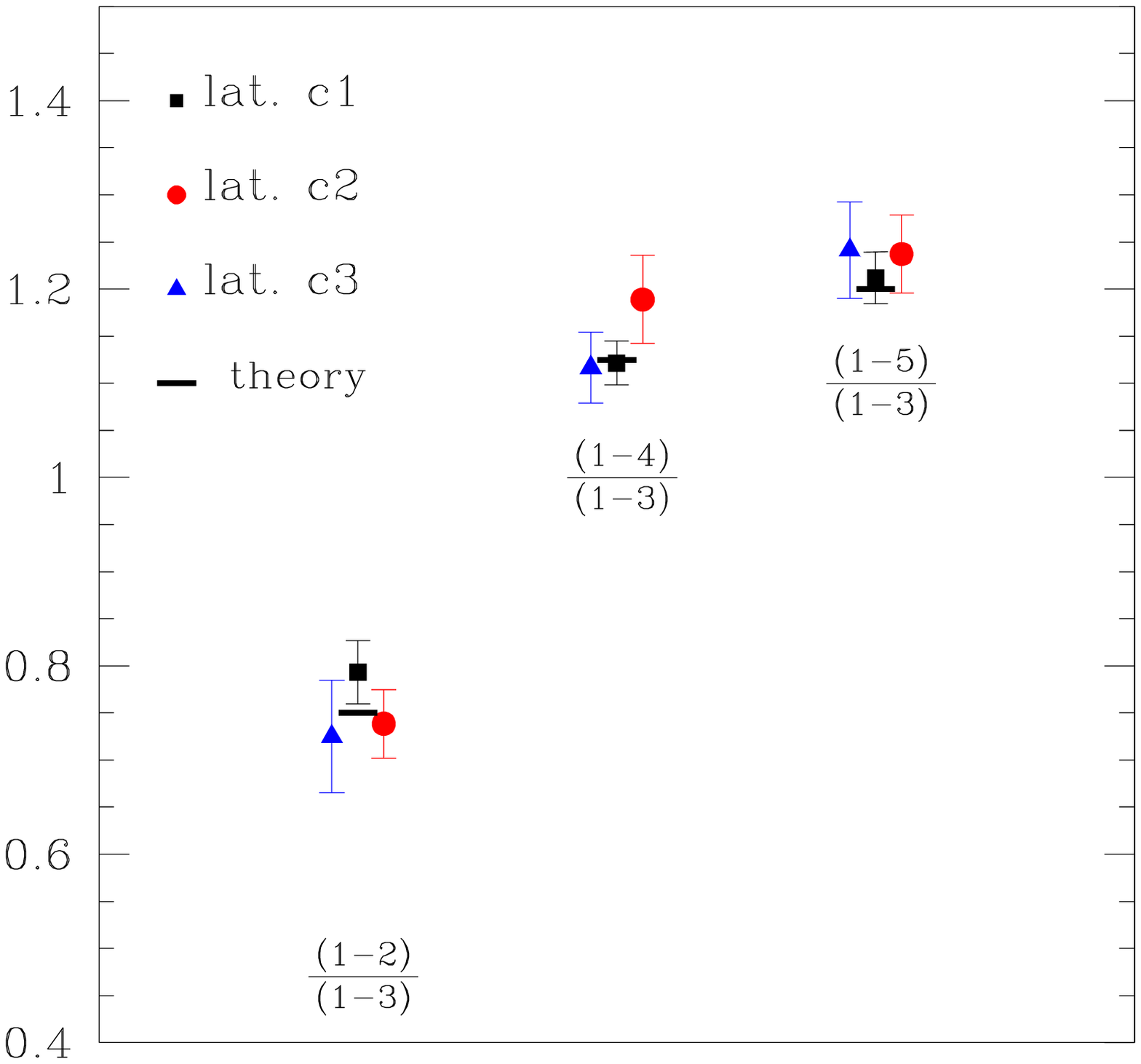}
\caption{On the left, the ratios ${(\chi_{\nu_1} - \chi_{\nu_2})}/{(\chi_{1} - \chi_{3})}$ as a 
function of the quark mass for the lattice c1 and 
for $(\nu_1,\nu_2)=(1,2),(1,4),(1,5)$. The points at $m=0$ represent the theoretical expectations 
from the Leutwyler-Smilga sum rule in Eq.~(\protect\ref{eq:ratios}). On the right the same ratios, 
at the lightest mass available for lattices c1, c2, c3, is compared with the 
theoretical expectations.}\label{fig:LS}
\end{figure}

\subsection{First Leutwyler-Smilga sum rule}
The sensitivity on the topology of the spectral density 
of the low-lying eigenvalues induces a non-trivial $\nu$-dependence 
in the chiral condensate. The latter has to be compared with 
the prediction of the chiral effective theory 
in Eq.~(\ref{eq:epsilon_l}). For the 
ratios
\be\label{eq:LS}
\frac{\chi_{\nu_1} - \chi_{\nu_2}}{\chi_{\nu_3} - \chi_{\nu_4}}
\ee
the prediction in the chiral limit is given 
in Eq.~(\ref{eq:ratios}). It is parameter-free and 
valid up to the NLO. For the lattice c1 and $\nu=1$--$5$, 
the ratios in Eq.~(\ref{eq:LS}) are shown in the first plot 
of Fig.~\ref{fig:LS} as 
a function of $(mV)a^{-3}$. As expected from Eq.~(\ref{eq:epsilon_l}) 
the mass dependence is very mild, and the two lightest points are 
consistent with a flat behaviour for all combinations. 
Results for lattices c2 and c3 show analogous features. 
In the plot on the right we report the points at the lightest mass
available for each of the three lattices, together with the theoretical expectations
from the first Leutwyler--Smilga sum rule in Eq.~(\ref{eq:ratios}). 
The very good agreement supports the 
fact that the topology dependence of these ratios is well 
reproduced in the (quenched) chiral effective theory. 
This is one of the main results of this paper.

\subsection{Discretization effects}\label{sec:la}
\begin{figure}[!t]
\begin{center}
\includegraphics[width=8cm]{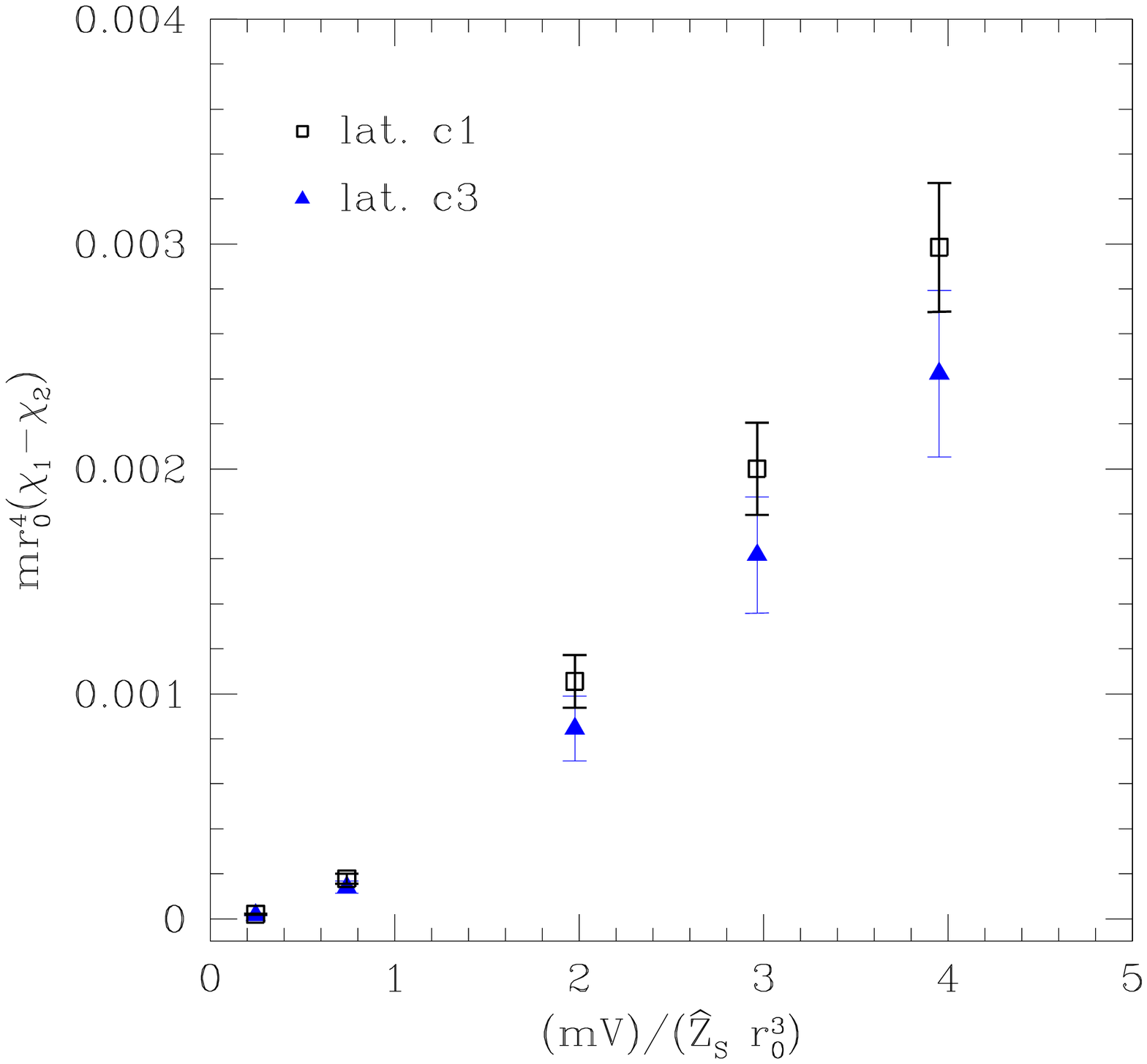}
\caption{The quantity $mr_0^4(\chi_{\nu_1}-\chi_{\nu_2})$ as a function of 
$mV/(\hat{Z}_Sr_0^3)$ for the lattices c1 and c3.}\label{fig:la}
\end{center}
\end{figure} 
Since we use fermions with an exact chiral symmetry, 
the leading discretization effects in our observables are of order $a^2$. 
We can estimate their magnitude by comparing renormalized dimensionless quantities 
calculated on the lattices c1 and c3. In Fig.~\ref{fig:la} we show the quantity 
$m r_0^4(\chi_{\nu_1}-\chi_{\nu_2})$ as a function of $mV/(\hat{Z}_S r_0^3)$. 
Within our statistical errors we do not observe significant 
deviations between the points of the two data sets. A similar behaviour is observed 
in the other topological sectors. Discretization effects are small and are likely to be 
comparable or below our statistical errors.
This is not surprising: quenched computations of 
various physical quantities carried out with Neuberger fermions show small 
discretization effects at the lattice spacings of our 
simulations~\cite{Babich:2005ay,Wennekers:2005wa}. A continuum limit extrapolation 
of our results is beyond the scope of this paper, and we find it of limited interest 
before removing the quenched approximation. 

\subsection{Extraction of the low-energy constant}
For every lattice and for $(\nu_1,\nu_2)=(0,1)$, $(1,2)$, $(2,3)$, we extract the 
value of the low-energy constant $\Sigma_{\rm eff}/Z_S$
from a one-parameter fit of the observables $\chi_{\nu_1}-\chi_{\nu_2}$,
with the infrared tail truncated as described in Sec.~\ref{sec:lma}, 
to the corresponding functional form predicted by the chiral effective 
theory. Combinations with  $\nu_1,\nu_2\ge 4$
have larger statistical errors and are not considered in 
this analysis.
%%%%%%%%%%%%%%%%%%%%%%%%%%%%%%%%%%%%%%%%%%%%%%%%%%%%%%%%%%%%%%%%%%%%%%%%%%%%%%%%%%%%%%%
\begin{figure}[ht]
\begin{center}
\includegraphics[width=8cm]{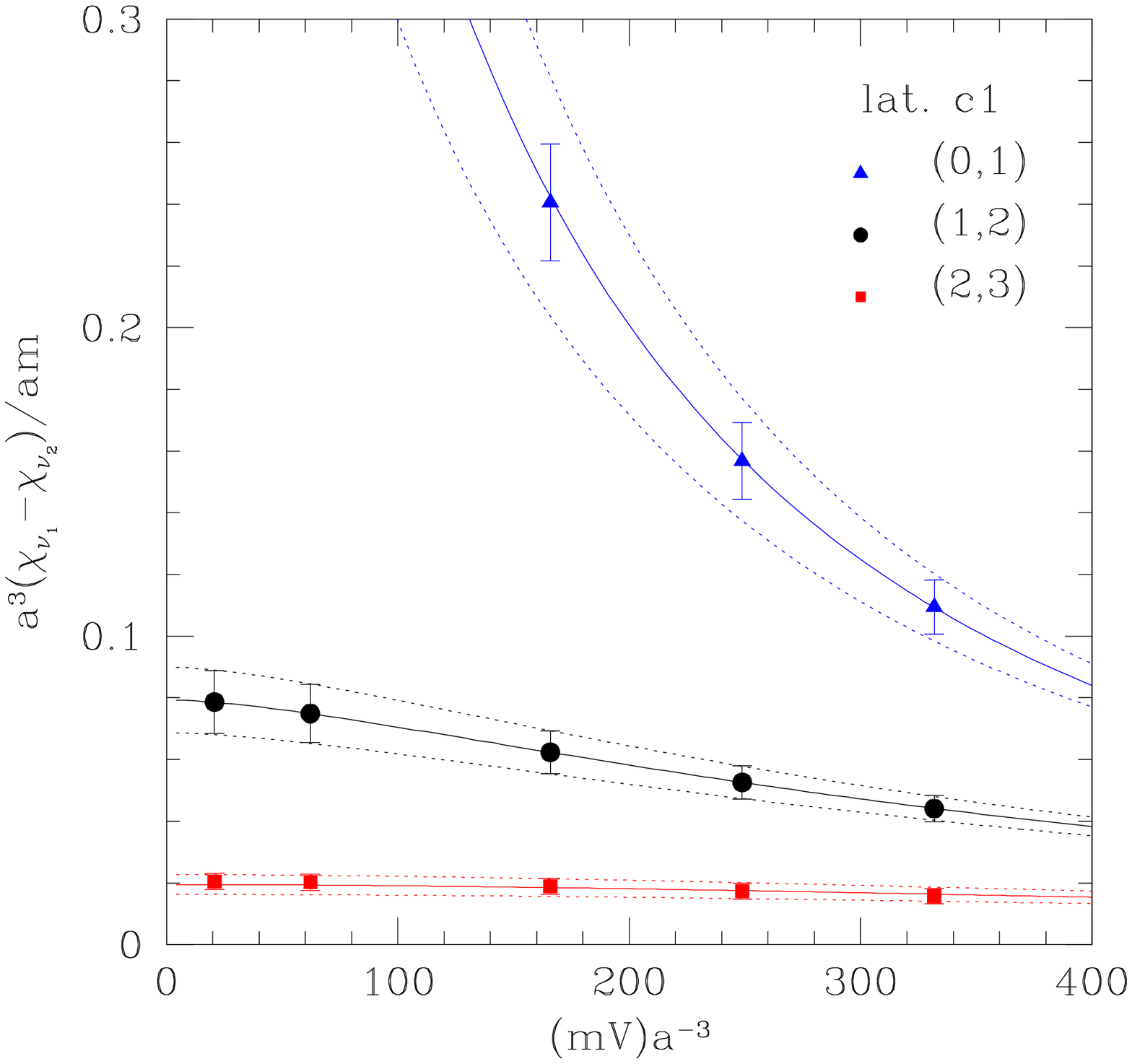}
\caption{The function $\tilde \chi_{\nu_1}-\tilde\chi_{\nu_2}$ is 
superimposed on data for $\chi_{\nu_1}-\chi_{\nu_2}$ from lattice c1.}\label{fig:fit}
\end{center}
\end{figure}
%%%%%%%%%%%%%%%%%%%%%%%%%%%%%%%%%%%%%%%%%%%%%%%%%%%%%%%%%%%%%%%%%%%%%%%%%%%%%%%%%%%%%%%
\begin{table}[!ht]
\begin{center}
\begin{tabular}{|c c c|}
\hline
lat    &  $\nu_1-\nu_2$  & $a^3 \Sigma_{\rm eff}/Z_S$ \\
\hline
c1     &  0-1            &  0.0040(6)                \\ 
       &  1-2            &  0.0039(3)           \\ 
       &  2-3            &  0.0034(3)           \\
\hline 
c2     &  0-1            &  0.0035(8)           \\
       &  1-2            &  0.0049(9)           \\
       &  2-3            &  0.0040(5)           \\
\hline
c3     &  0-1            &  0.0015(3)           \\
       &  1-2            &  0.00178(18)           \\
       &  2-3            &  0.00188(12)           \\
\hline
\end{tabular} 
\caption{Results for $a^3\Sigma_{\rm eff}/Z_S$ as determined from a 
fit of the lattice data (see text).}\label{tab:sigma_results}
\end{center}
\end{table}
%%%%%%%%%%%%%%%%%%%%%%%%%%%%%%%%%%%%%%%%%%%%%%%%%%%%%%%%%%%%%%%%%%%%%%%%%%%%%%%%%%%%%%%
The fit always reproduces the data very well, and the uncorrelated 
$\chi^2$/d.o.f. are typically very small. The results for $\Sigma_{\rm eff}/Z_S$
are reported in Tab.~\ref{tab:sigma_results}. 
At small values of $(mV)$ the quenched effective theory predicts 
a logarithmic behaviour of the form $(mV)\log{(mV)}$ in $\chi_0$, 
which is clearly seen in our data. If this term is removed from the fit
function, the latter is not compatible with the data anymore.
For illustration in Fig.~\ref{fig:fit} 
the function $\tilde \chi_{\nu_1}-\tilde\chi_{\nu_2}$ is superimposed on 
the data for $\chi_{\nu_1}-\chi_{\nu_2}$ from lattice c1.

The values of $a^3 \Sigma_{\rm eff}/Z_S$ are in good agreement 
within each data set, a consequence of the fact that the effective 
theory reproduces the topology dependence observed in the data. The 
results from lattices c1 and c2 confirm that, within our statistical errors, 
we do not observe a volume dependence in $\Sigma_{\rm eff}$. By taking  $\alpha=0$, 
the topological susceptibility from Ref.~\cite{DelDebbio:2004ns} and 
the quenched value of $F$ from Ref.~ \cite{Giusti:2004yp}, 
the NLO formula in Eq.~(\ref{eq:fvcor}) suggests for ${\Sigma_{\rm eff}(L\simeq \;2.0\;\rm{fm})}/
{\Sigma_{\rm eff}(L\simeq \;1.5\;\rm{fm})}$ a positive deviations from 1 of the order of
10 per cent. This is comparable to our statistical uncertainty, and the same 
holds if we vary $\alpha$ within a reasonable range of values 
\footnote{Explicitly, one 
finds  ${\Sigma_{\rm eff}(L= \;2.0\;\rm{fm})}/{\Sigma_{\rm eff}(L= \;1.5\;\rm{fm})}\sim
1+0.09+0.03\alpha$.}. Our best result for the renormalization group invariant condensate
from the lattices c1 and c3 is for the combination $(\nu_1,\nu_2)=(1,2)$, while from lattice c2 is for $(2,3)$:
\begin{eqnarray}
({\hat{\Sigma}_{\rm eff}r_0^3})(L=1.5\;\rm{fm})      & = & 0.33(3) \qquad \mathrm{c1}\, ,\\
({\hat{\Sigma}_{\rm eff}r_0^3})(L=2.0\;\rm{fm})      & = & 0.34(5) \qquad \mathrm{c2}\, ,\\
({\hat{\Sigma}_{\rm eff}r_0^3})(L=1.5\;\rm{fm})      & = & 0.29(3) 
\qquad \mathrm{c3}\, .
\end{eqnarray}
We take as our best estimate of the condensate at $L=1.5$~fm the result from 
lattice c3, which is the one with the finer lattice spacing. The 
latter, converted into the more usual $\overline{\rm MS}$-scheme at 2 GeV~\cite{Garden:1999fg}, is
\be\label{eq:f1}
r_0^3 \Sigma_{\rm eff}^{\overline{\rm MS}}(2\; \rm{GeV})=0.40(4)\;\; \mathrm{at} \;\;
 L=1.5~\mathrm{fm},
\ee
and if we use the phenomenological value $r_0= 0.5$~fm we obtain 
\be\label{eq:f2}
\Sigma_{\rm eff}^{\overline{\rm MS}}(2\; \rm{GeV})=\left(290\pm 11\;\rm{MeV}\right)^3 \;\; \mathrm{at} \;\; 
L = 1.5~\mathrm{fm}\; .
\ee
The errors in Eqs.~(\ref{eq:f1}) and (\ref{eq:f2}) do not include uncertainties due
to discretization effects. 
Our result is in the range expected from previous computations in the quenched 
approximation of QCD~\cite{Giusti:1998wy,Blum:2000kn,Hernandez:1999cu,Giusti:2001pk,Hernandez:2001hq,
DeGrand:2001ie,Hasenfratz:2002rp,Giusti:2003gf,Bietenholz:2006fj,Wennekers:2005wa}. The determinations 
from the infinite volume regime~\cite{Giusti:1998wy,Blum:2000kn,Giusti:2001pk,Hernandez:2001hq}
are affected by NLO chiral corrections which usually are not taken into account, and 
which are different from those in the finite volume regime. Even tough
the comparison cannot be very accurate, it is reassuring the good agreement 
of our result with some of these determinations. First exploratory studies of the chiral 
condensate in the finite volume regime were performed with small volumes and low 
statistics~\cite{Hernandez:1999cu,DeGrand:2001ie,Hasenfratz:2002rp}. 
This may be the reason why the large statistical 
fluctuations generated by the very small eigenvalues of 
the Dirac operator, which were first reported 
in Ref.~\cite{Hasenfratz:2002rp}, did not occur 
in some of these computations. 
A yet different method to extract $\Sigma_{\rm eff}$ was pursued in 
Ref.~\cite{Giusti:2003gf,Wennekers:2005wa,Bietenholz:2006fj}.
They compared the average value of the individual low-lying eigenvalues 
of the Dirac operator with the prediction of random matrix theory. 
Even if it relies on the non-trivial assumptions that the single eigenvalue
of the Dirac operator renormalizes with $1/Z_S$ and that the
random matrix theory reproduces the QCD spectrum in the infrared, the 
values of $\Sigma_{\rm eff}$ obtained in these computations are in good 
agreement with our result.

\section{Conclusions}\label{sec:conc}
The functional form of the (quenched) QCD 
chiral condensate in the volume range $5$--$16$~fm$^4$ 
and for masses $\hat m \Sigma V\leq 1$ turns out 
to be in good agreement, within our statistical errors, 
with the prediction of the chiral effective theory: 
the volume, mass and topology dependence are well 
reproduced by our data. Close to the chiral limit we observe 
a detailed agreement with the first Leutwyler--Smilga sum rule.
The low-energy constant $\Sigma_\mathrm{eff}$ is then extracted from 
a matching of the lattice results with the chiral formulas.

A technical progress which made the computation possible 
is the introduction of a numerical estimator which is stable
in the finite volume regime of (quenched) QCD.
Very small eigenvalues of the Dirac operator, which do occur
in the Monte Carlo history, do not generate large fluctuations, 
and a finite size scaling study of the condensate becomes 
feasible all the way down to the chiral limit. 
We have not tried to exploit the great freedom 
in the possible choices of the QCD observable, 
discussed in Sec.~\ref{sect_cond}, 
which are unambiguously defined in the continuum limit and 
could lead to a more precise determination of the low-energy 
constant $\Sigma_\mathrm{eff}$. This is beyond the 
scope of this paper and we leave it to future studies. 

This technique is directly applicable 
to the computation of the condensate in full QCD simulations.

\section*{Acknowledgments}
The present paper is part of an ongoing project whose final goal
is to extract low-energy parameters of (quenched) QCD from numerical 
simulations with GW fermions. We would like to thank P.~Hern\'andez, 
M.~L\"uscher, C. Pena,  P. Weisz, J. Wennekers and H. Wittig for many 
interesting  discussions and for allowing us to use part of the code 
developed together. Many thanks to M.~L\"uscher and M. Testa for interesting
discussions on the subject of this paper. Our calculations were performed 
on PC clusters at CILEA, at the University of Rome ``La Sapienza'' and at the 
University of Valencia. We thankfully acknowledge the computer resources 
and technical support provided by all these institutions and the University of 
Milano-Bicocca for its support. S.~N. is supported by the Marie Curie contract 
MEIF-CT-2006-025673 and was partially supported by EC Sixth Framework Program 
under the contract MRTN-CT-2006-035482 (FLAVIAnet).

\appendix
\section{Definitions and conventions}\label{appa}
We consider the Euclidean space-time discretized on a lattice with spacing $a$. 
Our conventions for the Dirac matrices are
\begin{equation}
\gamma_\mu^\dagger=\gamma_\mu\;\;\;\{\gamma_\mu,\gamma_\nu\}=2\delta_{\mu\nu}\; ,
\end{equation}
with 
\begin{equation}
\gamma_5=\gamma_0\gamma_1\gamma_2\gamma_3=\left(\begin{array}{cc}
1 &  0\\
0 &  -1 
\end{array}\right)\; . 
\end{equation}
and the chiral projectors defined as $P_{\pm}=(1\pm\gamma_5)/2$.
The massless Neuberger-Dirac operator is defined as 
\begin{equation}\label{eq:neu}
D=\frac{1}{\overline a}\left\{1+\gamma_5{\rm sign}({\cal Q})\right\}\; ,
\end{equation}
where
\begin{equation}
{\cal Q}=\gamma_5(aD_W-1-s)\; ,\;\;\;\;|s|<1\; ,\;\;\;\overline a=\frac{a}{1+s}\; ,
\end{equation}
and $D_W$ is the Wilson Dirac operator
\begin{equation}
D_W=\frac{1}{2}\left\{ \gamma_\mu(\nabla^*_\mu+\nabla_\mu)-a\nabla^*_\mu\nabla_\mu\right\}.
\end{equation}
The covariant forward and backward derivatives $\nabla_\mu$ and $\nabla^*_\mu$ 
are  
\begin{eqnarray}
\nabla_{\mu}\psi(x) & = & \frac{1}{a}\left\{ U(x,\mu)\psi(x+a\hat{\mu})-\psi(x)  \right\}\; ,\\
\nabla^*_\mu\psi(x) & = & \frac{1}{a}\left\{\psi(x)-  U(x-a\hat{\mu},\mu)^{-1}\psi(x-a\hat{\mu})   \right\},
\end{eqnarray}
where $U(x,\mu)\;\in$ SU(3) are the gauge variables and $\hat{\mu}$ is the versor in direction $\mu$. 
The massive quark propagator is 
\begin{equation}
%\bcontraction[0.75ex]{S_m(x,y)=}{\psi(x)}{}{\psi(x)} 
\bcontraction[0.75ex]{S_m(x,y)=}{\tilde \psi}{(x)}{\overline \psi} 
{S_m(x,y)=}{\tilde \psi (x)}{\overline \psi (y)} =
\frac{1}{1-\frac{\overline{a}m}{2}}\left\{D_m^{-1}  -\frac{\overline a}{2} \right\}(x,y)\; ,
\end{equation}
where
\begin{equation}\label{eq:psitilde}
\tilde{\psi}=\left(1-\frac{\overline a}{2}D  \right)\psi\; .
\end{equation}

\section{Renormalization of the scalar density}\label{appb}
The bare singlet scalar density 
\be
S^0(x) = \overline{\psi}(x)\tilde\psi(x)
\ee
needs to be renormalized to make finite its correlation 
functions with other gauge invariant operators
inserted at a physical distance. $S^0$ itself and 
the identity multiplied by proper combinations 
of the mass matrix are the only operators 
with dimension $d\leq 3$ and the same transformations properties 
under the $SU(N_f)_\mathrm{L}\otimes SU(N_f)_\mathrm{R}$ group. 
For quark masses degenerate and real, the renormalized operator 
can then be written as
\be
\hat S^0(x) = Z_S \left[S^0(x) + c_1 m + c_2 m^3\right]\; ,
\ee 
where $Z_S$ is the logarithmic-divergent renormalization constant of 
the scalar density, and $c_1$ and $c_2$ are subtraction coefficients
which at asymptotically large $a$ diverge as $1/a^2$ and $\ln (a)$ 
respectively\footnote{Notice that for $N_f=3$ the chiral group allows 
for a term proportional to $m^2$. Since the topological charge distribution 
is ultraviolet finite, this term is not ultraviolet divergent.}. 
The coefficients $c_1$ and $c_2$ can be fixed, for example, by requiring 
that in the infinite volume limit the condensate satisfies
\bea
\frac{d}{d \hat m}\; \langle \hat S^0 \rangle\Big|_{\hat m=0} & = & 0\; ,\\[0.25cm]
\frac{d^3}{d \hat m^3} \langle \hat S^0 \rangle\Big|_{\hat m=0} & = & 0\; .
\eea
The value of the renormalized condensate for $\hat m >0$ is therefore prescription
dependent, a well known fact in QCD.
The cumulants of the topological charge $Q$ are ultraviolet finite, and 
thus the distribution of the topological charge 
as defined in Eq.~(\ref{eq:Q})~\cite{Giusti:2001xh,Giusti:2004qd,Luscher:2004fu}.  
This implies that correlation functions of renormalized local operators 
inserted at a physical distance are finite also in the theory at fixed topology. 
In particular the combinations 
\be
\langle \hat S^0 \rangle_{\nu_1} - \langle \hat S^0 \rangle_{\nu_2}
\ee
are unambiguously defined outside the chiral limit, i.e. they are independent 
on the particular prescription chosen to renormalize the chiral 
condensate at finite mass.

%\bibliographystyle{h-elsevier}   
%\bibliography{lattice}        
\end{document}